\documentclass[a4paper,fleqn,usenatbib]{mnras}

\usepackage[T1]{fontenc}
\usepackage{ae,aecompl}

\usepackage{graphicx}	
\usepackage{amsmath}	
\usepackage{amssymb}	
\usepackage{multirow}
\usepackage{tabularx}
\usepackage{lscape}
\usepackage[usenames,dvipsnames]{xcolor}
\usepackage{float}
\usepackage{placeins}
\usepackage{color}

\title[The Sun's RV variations]{The Sun as a planet-host star: Proxies from SDO images for HARPS radial-velocity variations\thanks{Based on observations made with the HARPS instrument on the 3.6 m telescope under the program ID 088.C-0323 at Cerro La Silla (Chile), and the Helioseismic and Magnetic Imager onboard the Solar Dynamics Observatory. The HARPS observations, together with tables for the results presented in this paper are available in electronic format at: http://dx.doi.org/10.17630/bb43e6a3-72e0-464c-9fdd-fbe5d3e56a09. The SDO/HMI images can be downloaded from: http://jsoc.stanford.edu.}}

\author[R. D. Haywood et al.]{R. D. Haywood$^{1,2}$\thanks{E-mail:
rhaywood@cfa.harvard.edu}, A. Collier Cameron$^{1}$, Y. C. Unruh$^{3}$, C. Lovis$^{4}$,  A.F. Lanza$^{5}$, 
\newauthor
J. Llama$^{1, 6}$, M. Deleuil$^{7}$, 
R. Fares$^{1, 5}$,  M. Gillon$^{8}$, C. Moutou$^{7}$, F. Pepe$^{4}$,  
\newauthor
D. Pollacco$^{9}$, D. Queloz$^{4}$, and D. S\'{e}gransan$^{4}$\\
$^{1}$SUPA, School of Physics and Astronomy, University of St Andrews, St Andrews KY16 9SS, UK \\
$^{2}$Harvard-Smithsonian Center for Astrophysics, 60 Garden Street, Cambridge, MA 02138, USA \\
$^{3}$Astrophysics Group, Blackett Laboratory, Imperial College London, London SW7 2AZ, UK \\
$^{4}$Observatoire de Gen\`{e}ve, 51 Ch. des Maillettes, 1290 Sauverny, Switzerland \\
$^{5}$INAF-Osservatorio Astrofisico di Catania, via S. Sofia, 78 - 95123 Catania. Italy\\
$^{6}$Lowell Observatory, 1400 West Mars Hill Road, Flagstaff, AZ 86001, USA \\
$^{7}$Aix Marseille Universit\'e, CNRS, LAM (Laboratoire d'Astrophysique de Marseille) UMR 7326, 13388, Marseille, France \\
$^{8}$Institut d'Astrophysique et de G\'{e}ophysique, Universit\'{e} de Li\`{e}ge, All\'{e}e du 6 ao\^{u}t 17, Bat. B5C, 4000 Li\`{e}ge, Belgium\\
$^{9}$Department of Physics, University of Warwick, Coventry CV4 7AL, UK
}
\date{Accepted 2016 January 20. Received 2016 January 15; in original form 2015 August 27.}

\pubyear{2015}

\begin{document}
\label{firstpage}
\pagerange{\pageref{firstpage}--\pageref{lastpage}}
\maketitle

\begin{abstract}

The Sun is the only star whose surface can be directly resolved at high resolution, and therefore constitutes an excellent test case to explore the physical origin of stellar radial-velocity (RV) variability. We present HARPS observations of sunlight scattered off the bright asteroid 4/Vesta, from which we deduced the Sun's activity-driven RV variations. In parallel, the HMI instrument onboard the Solar Dynamics Observatory provided us with simultaneous high spatial resolution magnetograms, Dopplergrams, and continuum images of the Sun in the Fe {\sc I} 6173\AA~line. 
We determine the RV modulation arising from the suppression of granular blueshift in magnetised regions and the flux imbalance induced by dark spots and bright faculae. The rms velocity amplitudes of these contributions are 2.40~m~s$^{-1}$ and 0.41~m~s$^{-1}$, respectively, which
confirms that the inhibition of convection is the dominant source of activity-induced RV variations at play, in accordance with previous studies.
We find the Doppler imbalances of spot and plage regions to be only weakly anticorrelated. Lightcurves can thus only give incomplete predictions of convective blueshift suppression. We must instead seek proxies that track the plage coverage on the visible stellar hemisphere directly. The chromospheric flux index $R'_{HK}$ derived from the HARPS spectra performs poorly in this respect, possibly because of the differences in limb brightening/darkening in the chromosphere and photosphere. 
We also find that the activity-driven RV variations of the Sun are strongly correlated with its full-disc magnetic flux density, which may become a useful proxy for activity-related RV noise.
\end{abstract}




\section{Introduction}\label{intro}

The surface of a star is constantly bustling with magnetic activity. Starspots and faculae/plage\footnote{Plages are formed in the upper photosphere, chromosphere and upper layers of the stellar atmosphere \citep{Lean1997,Murdin:2002td}. They are not part of the lower photosphere, where the continuum absorption lines originate, from which the RV of a star is measured; however, plage regions do map closely to faculae and sunspots \citep{Hall:2008uga,Schrijver:2002ht}.} 
inhibit convective motions taking place at the stellar surface, thus suppressing part of the blueshift naturally resulting from granulation. In addition, dark starspots 
(and bright faculae, to a lesser extent)
coming in and out of view as the star rotates induce an imbalance between the redshifted and blueshifted halves of the star, which translates into an RV variation. Stellar activity, through the perturbation of photospheric convection, induces RV variations on much longer timescales of the order of years, in tune with their magnetic cycles (\emph{eg.} see \citet{2010A&A...511A..54S, 2011A&A...535A..55D, 2015arXiv151006446D}).

Short-term activity-induced RV variations are quasi-periodic: they are modulated by the star's rotation and change as active regions (starspots and faculae) emerge, evolve and disappear.
The amplitude of these variations is 1-2 m~s$^{-1}$ in ``quiet" stars \citep{2010ApJ...725..875I}, but they are often larger than this and can mimic or conceal the Doppler signatures of orbiting planets.
This has resulted in several false detections (see \citet{Queloz:2001be,Bonfils:2007cz,Huelamo:2008bk,Boisse:2009gl,Boisse:2011bw,Gregory:2011vg,Haywood:2014hsa,Santos:2014ho,Robertson:2014iz} and many others).

Understanding the RV signatures of stellar activity, especially those at the stellar rotation timescale, is essential to develop the next generation of more sophisticated activity models and further improve our ability to detect and characterise (super-)Earths and even small Neptunes in orbits of a few days to weeks. 
In particular, identifying informative and reliable proxies for the activity-driven RV variations is crucial.
\cite{Desort:2007dt} found that the traditional spectroscopic indicators (the bisector span (BIS) and full width at half-maximum (FWHM) of the cross-correlation profile) and photometric variations do not give enough information for slowly rotating, Sun-like stars (low $v\sin i$) to disentangle stellar activity signatures from the orbits of super-Earth-mass planets. 

The Sun is a unique test case as it is the only star whose surface can be resolved at high resolution, therefore allowing us to investigate directly the impact of magnetic features on RV observations. Early attempts to measure the RV of the integrated solar disc did not provide quantitative results about the individual activity features responsible for RV variability.
\cite{1986AdSpR...6...89J} measured integrated sunlight using a resonant scattering spectrometer and found that the presence of magnetically active regions on the solar disc led to variations of up to 15~m~s$^{-1}$. They also measured the disc-integrated magnetic flux but did not find any significant correlation with RV at the time due to insufficient precision.
At about the same time, \cite{Deming:1987jj} obtained spectra of integrated sunlight with an uncertainty level below 5~m~s$^{-1}$, enabling them to see the RV signature of supergranulation. The trend they observed over the 2-year period of their observations was consistent with suppression of convective blueshift from active regions on the solar surface.
A few years later, \cite{Deming:1994he} confirmed the findings of both \cite{1986AdSpR...6...89J} and \cite{Deming:1987jj}, only with a greater statistical significance. 
Not all studies were in agreement with each other, however; \cite{1993ApJ...403..801M} recorded spectra of sunlight scattered off the Moon over a 5-year period and found that any variations due to solar activity were smaller than 4~m~s$^{-1}$. 

More recently, \cite{Molaro:2010kq} obtained HARPS spectra of the large and bright asteroid Ceres to construct a wavelength atlas for the Sun. They found that these spectra of scattered sunlight provided precise disc-integrated solar RVs, and proposed using asteroid spectra to calibrate high precision spectrographs used for planet hunting, such as HIRES and HARPS (see a recent paper by \citet{Lanza2015}).

In parallel, significant discoveries were made towards a precise quantitative understanding of the RV impact of solar surface features.
 \cite{Lagrange:2010ija} and \cite{Meunier:2010hc} used a catalogue of sunspot numbers and sizes and MDI/SOHO magnetograms to simulate integrated-Sun spectra over a full solar cycle and deduce the impact of sunspots and faculae/plage on RV variations. 
The work of \cite{Meunier:2010hc} also relied on the amplitude of convection inhibition derived by  \citet{1990A&A...231..221B}, based on spatially-resolved magnetogram observations of plage and quiet regions on the Sun (\emph{i.e.}, independently of full-disc RV measurements).

Sunspot umbrae and penumbrae are cooler and therefore darker than the surrounding quiet photosphere, producing a flux deficit at the local rotational velocity. Faculae, which tend to be spatially associated with spot groups, are slightly brighter than the quiet photosphere, producing a spectral flux excess at the local rotational velocity. Spots have large contrasts and small area coverage, while faculae have lower contrasts but cover large areas.
We thus expect their rotational Doppler signals to be of opposite signs (due to the opposite sign of their flux) and of roughly similar amplitudes. However, facular emission also occurs independently of spot activity  in the more extended  solar magnetic network \citep{2001ApJ...555..462C}, so their contributions do not cancel out completely. \citet{Meunier:2010hc} found that the residual signal resulting from the rotational Doppler imbalance of both sunspots and faculae is comparable to that of the sunspots on their own.
 \citet{Lagrange:2010ija} estimated the rotational perturbation due to sunspot flux deficit to be of the order of 1~m~s$^{-1}$. 
In a complementary study, \citet{Meunier:2010hc} also investigated on the effect of sunspots and faculae on the suppression of convective blueshift by magnetic regions. Sunspots occupy a much smaller area than faculae, and as they are dark, they contribute little flux, so their impact on the convective blueshift is negligible. Facular suppression of granular blueshift, however, can lead to variations in RV of up to 8-10 m~s$^{-1}$ \citep{Meunier:2010hc}.
\citet{2010A&A...519A..66M} estimated the disc-integrated solar RV variations expected from the suppression of convective blueshift, directly from MDI/SOHO Dopplergrams and magnetograms (in the 6768 \AA~Ni {\sc I} line). Their reconstructed RV variations, over one magnetic cycle agree with the simulations of \citet{Meunier:2010hc}, establishing the suppression of convective blueshift by magnetic features as the dominant source of activity-induced RV variations. \citet{2010A&A...519A..66M} also found that the regions where the convective blueshift is most strongly attenuated correspond to the most magnetically active regions, as was expected.

Following the launch of the Solar Dynamics Observatory (SDO, \cite{Pesnell:2012ik}) in 2010, continuous observations of the solar surface brightness, velocity and magnetic fields have become available with image resolution finer than the photospheric granulation pattern. This allows us to probe the RV variations of the Sun in unprecedented detail. 
In the present paper, we deduce the activity-driven RV variations of the Sun based on HARPS observations of the bright asteroid Vesta (Section~\ref{harps}). In parallel, we use high spatial resolution continuum, Dopplergram and magnetogram images of the Fe {\sc I} 6173\AA~line from the Helioseismic and Magnetic Imager (HMI/SDO, \cite{Schou:2012cz}) to model the RV contributions from sunspots (and pores), faculae and granulation via inhibition of granular blueshift and flux blocking (Section~\ref{sdo}). This allows us to create a model which we test against the HARPS observations (Section~\ref{reproduce}). Finally, we discuss the implications of our study for the effectiveness of various proxy indicators for activity-driven RV variations in stars. We show that the disc-averaged magnetic flux could become an excellent proxy for activity-driven RV variations on other stars (Section~\ref{proxy}).

\begin{figure}
\centering
\includegraphics[width=0.25\textwidth]{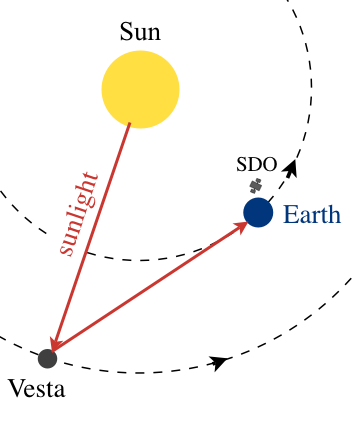}
\centering
\caption{Schematic representation of the Sun, Vesta and Earth configuration during the period of observations (not to scale).}
\label{geometry}
\end{figure}

\section{HARPS observations of sunlight scattered off Vesta}\label{harps}
\subsection{HARPS spectra}
The HARPS spectrograph, mounted on the ESO 3.6\,m telescope at La Silla was used to observe sunlight scattered from the bright asteroid 4/Vesta (its average magnitude during the run was 7.6). Two to three measurements per night were made with simultaneous Thorium exposures for a total of 98 observations, spread over 37 nights between 2011 September 29 and December 7. The geometric configuration of the Sun and Vesta relative to the observer is illustrated in Figure~\ref{geometry}.
At the time of the observations, the Sun was just over three years into its 11-year magnetic cycle; the SDO data confirm that the Sun showed high levels of activity. 

The spectra were reprocessed using the HARPS DRS pipeline \citep{Baranne:1996tb,Lovis:2007cp}. Instead of applying a conventional barycentric correction, the wavelength scale of the calibrated spectra was adjusted to correct for the 
 Doppler shifts due to the relative motion of the Sun and Vesta, and the relative motion of Vesta and the observer (see Section~\ref{relativistic}). 
The FWHM and BIS (as defined in \citet{Queloz:2001be}) and $\log R'_{\rm HK}$ index were also derived by the pipeline.
The median, minimum and maximum signal to noise ratio of the reprocessed HARPS spectra at central wavelength 556.50 nm are 161.3, 56.3 and 257.0, respectively.
Overall, HARPS achieved a precision of 75 $\pm$ 25 cm~s$^{-1}$ (see Table~A1). The reprocessed HARPS cross-correlation functions and derived RV measurements are available in the Supplementary Files that are available online.

We account for the RV modulation induced by Vesta's rotation in Section~\ref{axial}, and investigate sources of intra-night RV variations in Section~\ref{instrument}. We selected the SDO images in such a way as to compensate for the different viewing points of Vesta and the SDO spacecraft: Vesta was trailing SDO, as shown in Figure~\ref{geometry}. This is taken into account in Section~\ref{timelag}.

\begin{figure}
\centering
\includegraphics[width=0.48\textwidth]{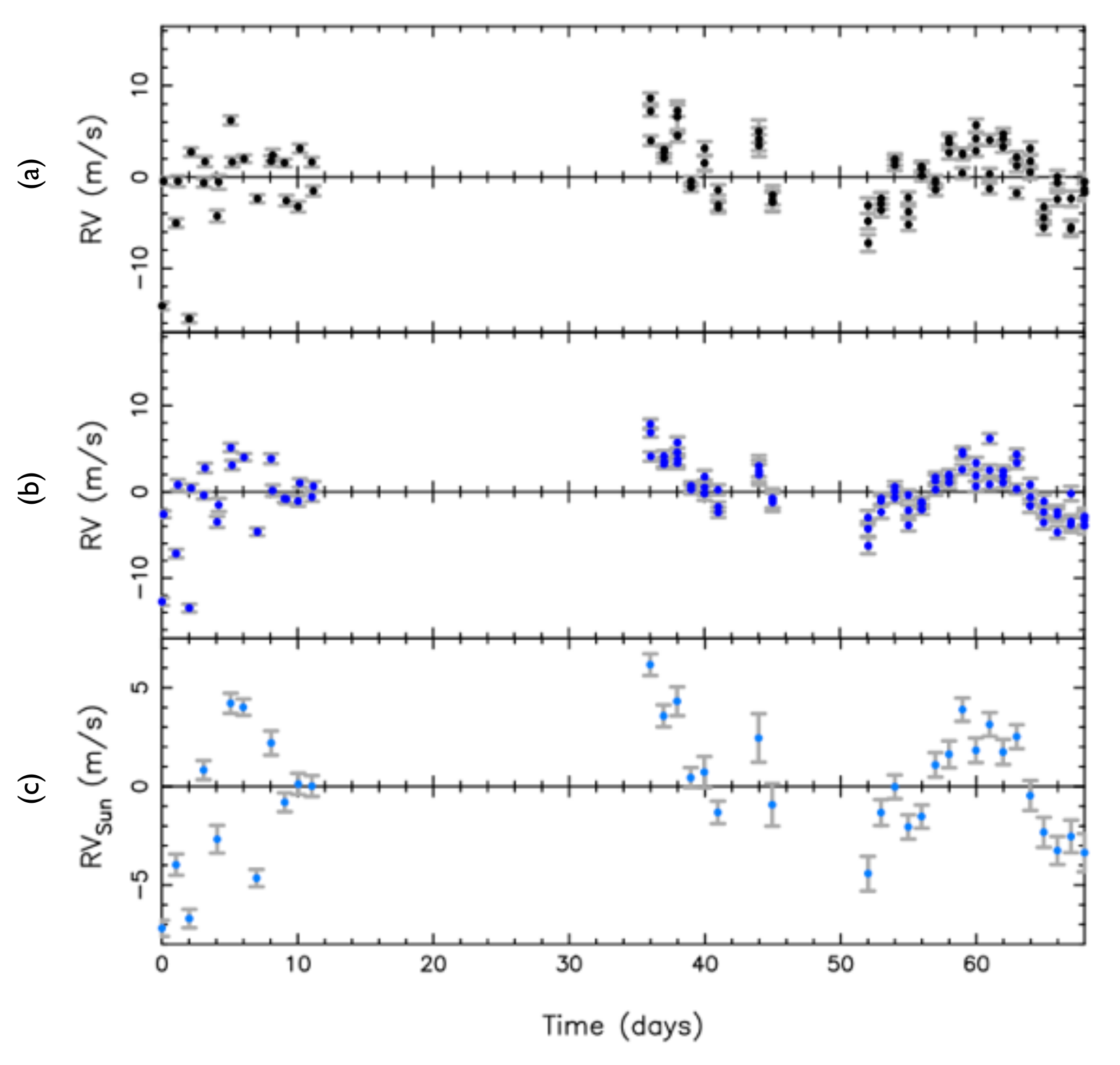}
\centering
\caption{\emph{Panel (a):} HARPS RV variations in the solar rest-frame, corrected for relativistic Doppler effects (but not yet corrected for Vesta's axial rotation). \emph{Panel (b):} HARPS RV variations of the Sun as-a-star (after removing the RV contribution of Vesta's axial rotation). \emph{Panel (c):} Nightly binned HARPS RV variations of the Sun as-a-star -- note the change in scale on the $y$-axis. The values for each time series are provided in the Supplementary Files that are available online.}
\label{two}
\end{figure}

\subsection{Solar rest frame} \label{rdot}
The data reduction pipeline for HARPS assumes that the target observed is a distant point-like star, and returns its RV relative to the solar system barycenter. In order to place the observed RVs of Vesta in the solar rest frame, we perform the following operation:
\begin{equation}
RV = RV_{\rm bary, Earth} + v_{\rm sv} + v_{\rm ve},
\end{equation}
where $RV_{\rm bary, Earth}$ is the barycentric RV of the Earth, \emph{i.e.} the component of the observer's velocity relative to the solar system barycentre, toward the apparent position of Vesta. It can be found in the fits header for each observation.
The two components $v_{\rm sv}$ and $v_{\rm ve}$, retrieved from the JPL {\sc horizons} database\footnote{Solar System Dynamics Group, Horizons On-Line Ephemeris System, 4800 Oak Grove Drive, Jet Propulsion Laboratory, Pasadena, CA  91109 USA -- Information: http://ssd.jpl.nasa.gov/, Jon.Giorgini@jpl.nasa.gov} correspond to:
\begin{description}
\item[-] $v_{\rm sv}$: the velocity of Vesta relative to the Sun at the instant that light received at Vesta was emitted by the Sun;
\item[-] $v_{\rm ve}$: the velocity of Vesta relative to Earth at the instant that light received by HARPS was emitted at Vesta.
\end{description}
This correction accounts for the RV contribution of all bodies in the solar system and places the Sun in its rest frame. The values of $RV_{\rm bary, Earth}$, $v_{\rm sv}$ and $v_{\rm ve}$ are given in the Supplementary Files that are available online.

\subsection{Relativistic Doppler effects}\label{relativistic}
The only relativistic corrections made by JPL {\sc horizons} are for gravitational bending of the light and relativistic aberration due to the motion of the observer (Giorgini, priv. comm.). We therefore must correct for the relativistic Doppler shifts incurred. 
The wavelength correction factor to be applied is given by \citet{Lindegren:2003in} as:
\begin{equation}
\frac{\lambda_{\rm e}}{\lambda_{\rm o}} 
= \frac{\sqrt{1 - \frac{v^{2}}{c^{2}}} }{1+ \frac{v \cos \theta_{\rm o}}{\rm c}},
\label{Doppler}
\end{equation}
where $\lambda_{\rm e}$ is the wavelength of the light at emission, $\lambda_{\rm o}$ is the wavelength that is seen when it reaches the observer, and $\theta_{\rm o}$ is the angle between the direction of the emitter at emission and the observed direction of the light at reception. $v$ is the total magnitude of the velocity vector of the observer relative to the emitter, and $c$ is the speed of light. 

We apply this correction twice:
\begin{itemize}
\item[-] The light is emitted by the Sun and received at Vesta. In this case, $v$ is the magnitude of the velocity of Vesta with respect to the Sun (decreasing from approximately 19.9 to 19.4 km.s$^{-1}$ over the duration of the HARPS run), and the radial component $v \cos \theta_{\rm o}$ is equal to $v_{\rm sv}$ (defined in Section~\ref{rdot}, starting at about 1.66 km.s$^{-1}$ and reaching 1.72 km.s$^{-1}$ at opposition near the middle of the run).
\item[-] Scattered sunlight is emitted from Vesta and received at La Silla. $v$ is the magnitude of the velocity of Vesta with respect to an observer at La Silla (increasing from 20 to 36 km.s$^{-1}$ over the run), and $v \cos \theta_{\rm o}$ is $v_{\rm ve}$ (starting at 19.8 km.s$^{-1}$ and reaching 23.3 km.s$^{-1}$ at opposition).
\end{itemize}

The two wavelength correction factors are then multiplied together in order to compute the total relativistic correction factor to be applied to the pixel wavelengths in the HARPS spectra, from which we derive the correct RVs, shown in Figure~\ref{two}~(a) (see also column 2 of Table~A1). 
All velocities are measured at the flux-weighted mid-exposure times of observation (MJD$_{\rm mid}$\_UTC).

The reader may also refer to Appendix A of \citet{Lanza2015} for further details on these Doppler corrections. All quantities used to calculate these effects are listed in the Supplementary Files that are available online.



\subsection{Sources of intra-night RV variations}

\subsubsection{Vesta's axial rotation}\label{axial}
Vesta rotates every 5.34 hours \citep{1951ApJ...114..500S}, so any significant inhomogeneities in its shape or surface albedo will induce an RV modulation. Vesta's shape is close to a spheroid \citep{Thomas:1997eh}, and \citet{2015ExA....39..461L} found that the RV modulation expected from shape inhomogeneities should not exceed 0.060~m~s$^{-1}$. 

\cite{1951ApJ...114..500S} presented a photometric study of the asteroid, and suggested that its surface brightness is uneven. He reported brightness variations $\delta m$ = 0.12 mag. 
To make a rough estimate of the amplitude of the RV modulation, we can assume that the brightness variations are due to a single dark equatorial spot on the surface of Vesta, blocking a fraction $\delta f$ of the flux $f$. $\delta m$ and $\delta f$ are related as follows: 
\begin{equation}
\delta m = \frac{2.5\, d(\ln f)}{\log (e)} \sim 1.08\, \frac{\delta f}{f},
\end{equation}

The fractional flux deficit caused by a dark spot can thus be approximated as:
\begin{equation}
\frac{\delta f}{f} \sim \delta m / 1.08 \sim 0.11.
\end{equation}

When the dayside of Vesta is viewed fully illuminated, this spot will give an RV modulation equal to:
\begin{equation}\label{aa}
\Delta RV_{\rm vesta} = - \frac{\delta f}{f} \,
v_{\rm eq} \,
\cos \theta \,
\sin \theta, 
\end{equation}
where $\theta$ is the angle between the spot on the asteroid and our line of sight, and increases from $-\pi/2$ to $+\pi/2$ as it traverses the visible daylight hemisphere. Due to foreshortening, the RV contribution is decreased by a factor $\cos \theta$. The line-of-sight velocity varies with $\sin \theta$.
The asteroid's equatorial velocity $v_{\rm eq}$ is given by:
\begin{equation}
v_{\rm eq} = 2\pi\, \frac{R_{\rm vesta}}{P_{\rm rot}}.
\end{equation}
Using a mean radius $R_{\rm vesta}~=~262.7$ km \citep{2012Sci...336..684R} and the rotational period $P_{\rm rot}~=~5.34$ hours, we obtain $v_{\rm eq}~=~85.8$~m~s$^{-1}$. 
The maximum RV amplitude of Vesta's rotational modulation, expected at $\theta~=~\pi/4$ is thus approximately 4.7~m~s$^{-1}$. The RV modulation due to surface brightness inhomogeneity should therefore dominate strongly over shape effects.


We find that this RV contribution is well modelled as a sum of Fourier components modulated by Vesta's rotation period:
\begin{equation}
\begin {aligned}\label{eqn6}
\Delta RV_{\rm vesta}(t) = C \cos(2\pi - \lambda(t)) + S \sin(2\pi - \lambda(t)),
 \end{aligned}
\end{equation}
where $\lambda(t)$ is the apparent planetographic longitude of Vesta at the flux-weighted mid-times of the HARPS observations and can be retrieved via the JPL {\sc horizons} database (the values of $\lambda$ are listed in Table~A1). $C$ and $S$ are scaling parameters, which we determine via a global optimal scaling procedure, performed with the solar activity contributions (see Section~\ref{model}). 
Since the phase-folded lightcurve of Vesta shows a double-humped structure \citep{1951ApJ...114..500S}, we also tested adding further Fourier terms modulated by the first harmonic of the asteroid's rotation. The improvement to the fit was negligible, so we preferred the simpler model of Equation~\ref{eqn6}.

Figure~\ref{two} (b) shows the RV observations obtained after subtracting Vesta's rotational signature, with coefficients $C$ and $S$ of Equation~\ref{eqn6} derived from the global fit of Section~\ref{model}. The night-to-night scatter has been reduced, even though much of it remains in the first block of observations; this is discussed in the following Section.

\subsubsection{Additional intra-night scatter in first half of HARPS run}\label{instrument}
The RV variations in the first part of the HARPS run (nights 0 to 11 in Figure~\ref{two}) contain some significant scatter, even after accounting for Vesta's rotation. 
This intra-night scatter does not show in the solar FWHM, BIS or $\log(R'_{\rm HK})$ variations. 
We investigated the cause of this phenomenon and excluded changes in colour of the asteroid or instrumental effects as a potential source of additional noise. 
Vesta was very bright (7.6 mag), so we deem the phase and proximity of the Moon unlikely to be responsible for the additional scatter observed. 

Solar p-mode oscillations dominate the Sun's power spectrum at a timescale of about 5 minutes. Most of the RV oscillations induced by p-mode acoustic waves are therefore averaged out within the 15-minute HARPS exposures. 
Granulation motions result in RV signals of several m~s$^{-1}$, over timescales ranging from about 15 minutes to several hours. Taking multiple exposures each night and averaging them together (as plotted in panel (c) of Figure~\ref{two}) can help to significantly reduce granulation-induced RV variations.

The velocity measurements are sensitive to any displacement of the image of Vesta from the centre of the 1-arcsec fibre. Vesta had a finite angular diameter of 0.49 arcsec at the start of the run, and 0.32 arcsec at the end. Light reflected from Vesta is blueshifted by 172 m~s$^{-1}$ at the approaching limb and redshifted by 172 m~s$^{-1}$ at the receding limb. We simulated the effects of seeing on an ensemble of photons originating from different points on Vesta's disc, applying random angular deviations with a gaussian seeing distribution. For a given displacement of Vesta from the centre of the fibre we computed the mean rotational velocity displacement of those photons falling within the fibre. We found an approximately linear dependence of the mean rotational velocity on mean offset from the centre of the fibre. Our numerical simulations suggest that a mean displacement by 0.1 arcsec of the image from the centre of the fibre in the direction orthogonal to Vesta's rotation axis, averaged over the exposure, gives a velocity error that is closely approximated by the empirical expression:

\begin{equation}
 \Delta v = 4.0\, (\frac{\theta}{0.49''}) \,/\, \sqrt{\frac{\sigma_{\rm seeing}}{1.0''}} \,{\rm m~s}^{-1},
\end{equation}
 where $\sigma_{\rm seeing}$ represents the full width at half-maximum of the gaussian seeing distribution.
This is of the same order as the excess RV scatter observed during the first part of the run, when successive observations within a night were widely separated at different airmasses. We conclude that small airmass-dependent guiding errors provide a plausible explanation. During the latter part of the run, the observations were contiguous and were thus conducted at similar airmass, leading to more consistent guiding and smaller scatter. Other factors such as asymmetric image quality, coma of the telescope varying with elevation (and sky rotation on an equatorial telescope), atmospheric dispersion, \emph{etc.} may also contribute to this effect.

%
%
%
The remaining variations, of order 7-10~m~s$^{-1}$, are modulated by the Sun's rotation and are caused by the presence of magnetic surface markers, such as sunspots and faculae. These variations are the primary focus of this paper, and we model them using SDO/HMI data in Section~\ref{sdo}.


\subsection{Time lag between Vesta and SDO observations}\label{timelag} 
At the time of the observations, the asteroid Vesta was trailing the SDO spacecraft, which orbits the Earth  (see Figure~\ref{geometry}).
In order to model the solar hemisphere facing Vesta at time $t$, we used SDO images recorded at $t+ \Delta t$, where $\Delta t$ is proportional to the difference in the Carrington longitudes of the Earth/SDO and Vesta at the time of the HARPS observation. These longitudes were retrieved from the JPL {\sc horizons} database.
The shortest delay, at the start of the observations was $\sim$ 2.8 days, while at the end of the observations it reached just over 6.5 days (see Table~A1). We cannot account for the evolution of the Sun's surface features during this time, and must assume that they remain frozen in this interval. 
The emergence of sunspots can take place over a few days, but in general large magnetic features (sunspots and networks of faculae) evolve over timescales of weeks rather than days.  Visual inspection of animated sequences of SDO images obtained during the campaign revealed no major flux emergence events on the visible solar hemisphere.

\section{Pixel statistics from SDO/HMI images}\label{sdo}
In the second part of this analysis we aim to determine the RV contribution from granulation, sunspots and facular regions. We use high-resolution full-disc continuum intensity, line-of-sight Doppler images and line-of-sight magnetograms from the HMI instrument (Helioseismic and Magnetic Imager) onboard SDO\footnote{HMI data products can be downloaded online via the Joint Science Operations Center website: http://jsoc.stanford.edu.}. These were retrieved for the period spanning the HARPS observations of Vesta at times determined by the time lags detailed in Section~\ref{timelag} (the exact date and time stamps of the images are listed in the Supplementary Files that are available online). 
SDO/HMI images the solar disc in the Fe {\sc I} 6173\AA~line at a cadence of 45 seconds, with a spatial resolution of 1'' using a CCD of 4096$\times$4096 square pixels. 
We first converted the SDO/HMI images from pixel coordinates to heliographic coordinates, \emph{i.e.} to a coordinate system centered on the Sun. This coordinate system is fixed with respect to the Sun's surface and rotates in the sidereal frame once every 25.38 days, which corresponds to a Carrington rotation period \citep{1859MNRAS..19...81C}. 
A surface element on the Sun, whose image falls on pixel $ij$ of the instrument detector, is at position ($w_{ij}, n_{ij}, r_{ij}$) relative to the centre of the Sun, where $w$ is westward, $n$ is northward and $r$ is in the radial direction (see \cite{Thompson:2006gl} for more details on the coordinate system used). The spacecraft is at position (0, 0, $r_{\rm sc}$).
The $w, n, r$ components of the spacecraft's position relative to each element $ij$ can thus be written as:
\begin{equation}
\begin{aligned}
\delta w_{ij} = w_{ij} - 0 \\
\delta n_{ij} = n_{ij} - 0 \\
\delta r_{ij} = r_{ij} - r_{\rm sc}
\end{aligned}
\end{equation}

%
%

The spacecraft's motion and the rotation of the Sun introduce velocity perturbations, which we determine in Sections~\ref{spacecraft} and~\ref{sunrot}, respectively. These two contributions are then subtracted from each Doppler image, thus revealing the Sun's magnetic activity velocity signatures. We compute the RV variations due to the suppression of convective blueshift and the flux blocked by sunspots on the rotating Sun in Sections~\ref{rvc} and~\ref{rvr}. We show that the Sun's activity-driven RV variations are well reproduced by a scaled sum of these two contributions in Section~\ref{scaling}. Finally, we compute the disc-averaged magnetic flux and compare it as an RV proxy against the traditional spectroscopic activity indicators in Section~\ref{proxy}.

\begin{table}
\centering
\caption{Solar differential rotation profile parameters from Snodgrass \& Ulrich (1990).}
 \begin{tabular}{@{}llrrrrlrlr@{}}
 Parameter & Value (deg day$^{-1}$) \\
 \hline
$\alpha_1$ & 14.713 \\
$\alpha_2$ & -2.396 \\
$\alpha_3$ & -1.787 \\
\hline
\label{snodgrass}
\end{tabular}
\end{table}

\subsection{Spacecraft motion} \label{spacecraft}
The $w,~n,~r$ components of the velocity incurred by the motion of the spacecraft relative to the Sun, ${v_{{\rm sc}}}$, are given in the fits header of each SDO/HMI observation to a precision of $10^{-6}$~ms$^{-1}$.
The magnitude of the spacecraft's velocity away from pixel $ij$ can therefore be expressed as:
\begin{equation}
v_{{\rm sc}, ij} = - \frac{\delta w_{ij} \, v_{{\rm sc},w_{ij}} + \delta n_{ij} \, v_{{\rm sc},n_{ij}} + \delta r_{ij} \, v_{{\rm sc}, r_{ij}}}
{d_{ij}},
\end{equation}
where:
\begin{equation}
d_{ij} = \sqrt{\delta w_{ij}^2 + \delta n_{ij}^2 + \delta r_{ij}^2}
\end{equation}
is the distance between pixel $ij$ and the spacecraft. We note that all relative velocities in this paper follow the natural sign convention that velocity is rate of change of distance.

\begin{figure*}
\centering
\includegraphics[width =0.8\textwidth]{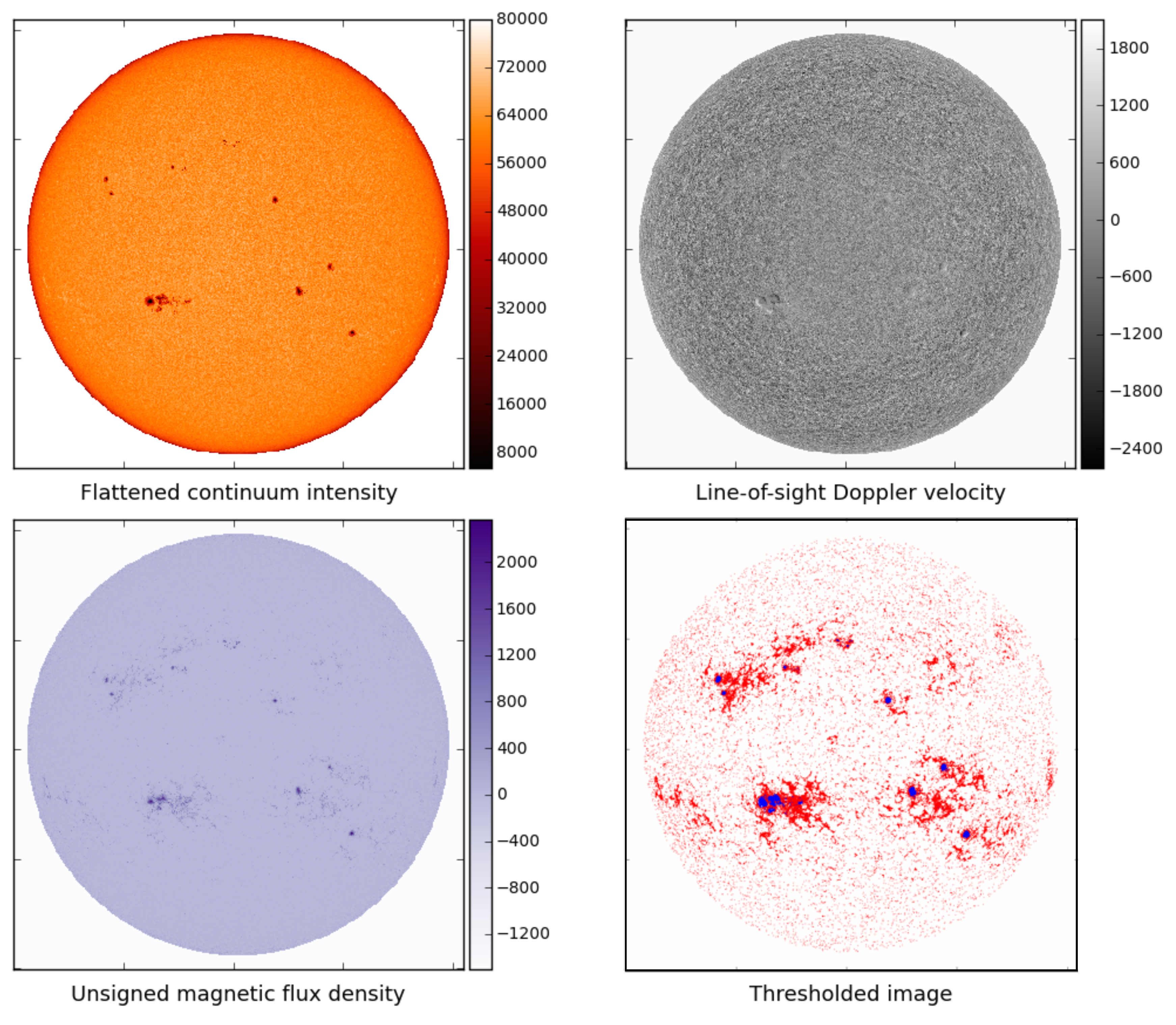}
\centering
\caption{\emph{First three panels:} SDO/HMI flattened intensity $I_{\rm flat}$, line-of sight velocity $v$ (km~s$^{-1}$) for the non-rotating Sun, and unsigned radial magnetic flux $|B_{\rm r}|$ (G) of the Sun, observed on 2011, November 10 at 00:01:30 UTC. \emph{Last panel:} our thresholded image, highlighting faculae (red/lighter shade pixels) and sunspots (blue/darker shade pixels). For this set of observations (representative of the whole run), faculae account for 9\% of the total pixel count, while sunspots account for less than 0.4\%. The remaining $\sim$ 90\% of the pixels on the solar disc are magnetically quiet.}
\label{pictures}
\end{figure*}

\begin{figure}
\centering
\includegraphics[width = 0.5\textwidth]{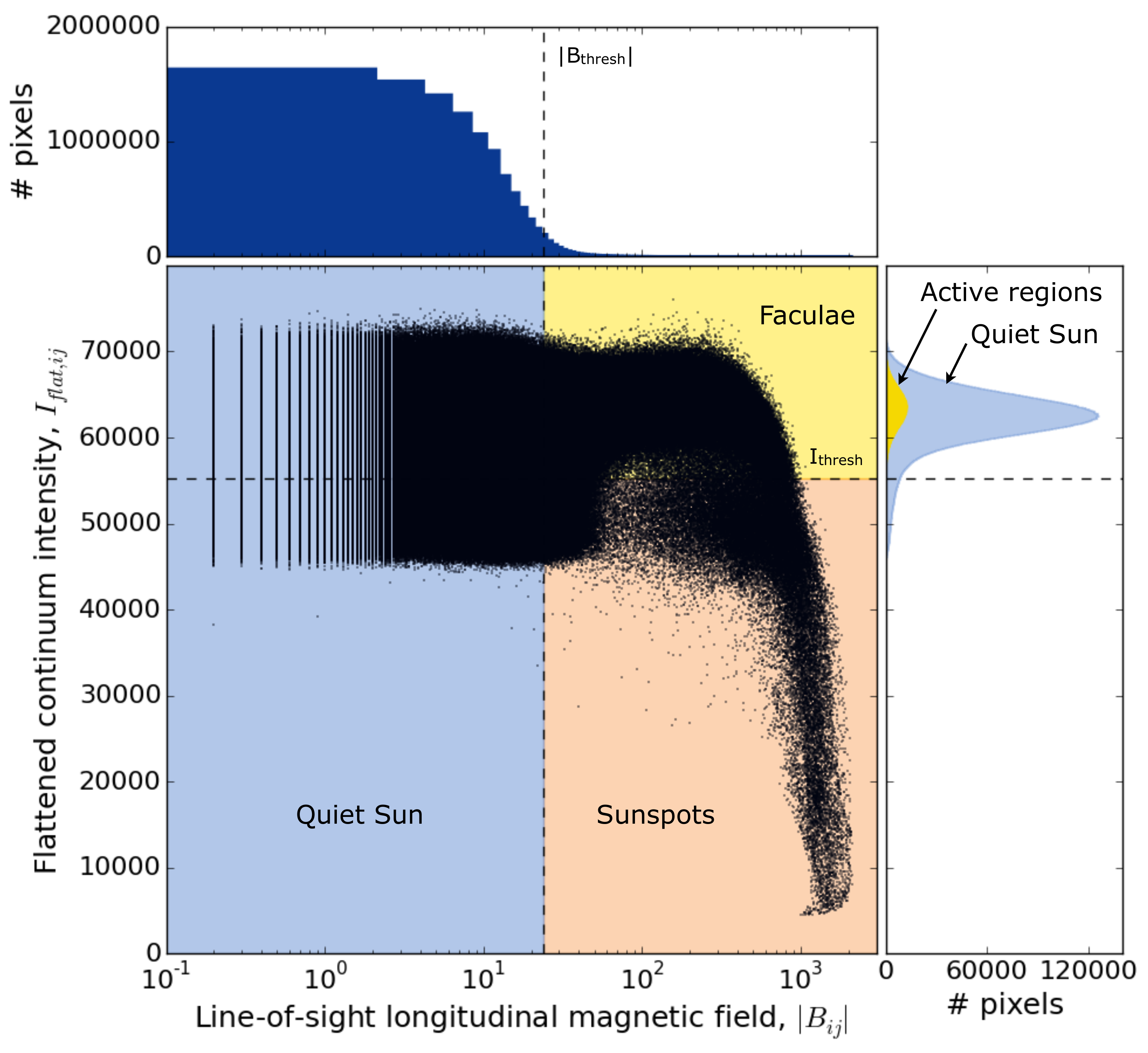}
\centering
\caption{Observed pixel line-of-sight (unsigned) magnetic field strength, $|B_{{\rm obs},ij}|$ (G), as a function of flattened intensity $I_{{\rm flat}, ij}$, for the Sun on 2011, November 10 at 00:01:30 UTC. The top and right histograms show the distributions of $|B_{{\rm obs},ij}|$ and $I_{{\rm flat}, ij}$, respectively. The dashed lines represent the cutoff criteria selected to define the quiet photosphere, faculae and sunspots.}
\label{patchwork}
\end{figure}

\subsection{Solar rotation}\label{sunrot}
The solar rotation as a function of latitude was measured by \cite{Snodgrass:1990ez} in low resolution full-disc Dopplergrams and magnetograms obtained at the Mount Wilson 150 foot tower telescope between 1967 and 1987. By cross-correlating time series of Dopplergrams and magnetograms, they were able to determine the rate of motion of surface features (primarily supergranulation cells and sunspots) and deduce the rate of rotation of the Sun's surface as a function of latitude. 
The solar differential rotation profile $\omega(\phi)$ at each latitude $\phi$ is commonly described by a least squares polynomial of the form:
\begin{equation}
\omega(\phi) = \alpha_1 + \alpha_2 \sin^2\phi + \alpha_3 \sin^4\phi.
\end{equation}
The best fit parameters found by Snodgrass \& Ulrich (1990), used in this analysis, are given in Table~\ref{snodgrass}. We apply this rotation profile in the heliographic frame to determine the $w, n, r$ components induced by the solar rotation velocity along the line of sight to a given image pixel, $v_{{\rm rot}, w}$, $v_{{\rm rot}, n}$ and $v_{{\rm rot}, r}$. Normalising again by $d$, we can write:
\begin{equation}
v_{{\rm rot}} = - \frac{\delta w \, v_{{\rm rot},w} + \delta n \, v_{{\rm rot},n} + \delta r \, v_{{\rm rot}, r}}
{d}.
\end{equation}

\subsection{Flattened continuum intensity}
We flatten the continuum intensity images using a fifth order polynomial function $L_{ij}$ with the limb darkening constants given in Astrophysical Quantities \citep{Allen:1973us}, through the IDL subroutine \emph{darklimb\_correct.pro}\footnote{Source code available at: \\ http://hesperia.gsfc.nasa.gov/ssw/gen/idl/solar/}.
The flattened and non-flattened continuuum intensities are related via the limb-darkening function $L$ as follows:
\begin{equation}
I_{{\rm flat}, ij} = \frac{I_{ij}}{L_{ij}}.
\end{equation}\label{limb}

We also define a scaling factor $\hat{K}$ which will be applied to $L_{ij}$ later on:
\begin{equation}
\hat{K} = \frac{\sum_{ij} I_{ij} \, L_{ij} \, W_{ij}}{\sum_{ij} L^{2}_{ij} \, W_{ij}},
\end{equation}
where the weighting factor $W_{ij}$ is set to unity for quiet-Sun pixels, and zero elsewhere.

\subsection{Unsigned magnetic field strength}\label{magnetic}
The SDO/HMI instrument measures the line-of-sight magnetic field strength $B_{\rm obs}$. 
The magnetic field of the Sun stands radially out of the photosphere with a strength $B_{\rm r}$. 
Due to foreshortening, the observed field $B_{\rm obs}$ is less than the true (radial) field by a factor:
\begin{equation}
\mu_{ij} = \cos \theta_{ij},
\end{equation}
where $\theta_{ij}$ is the angle between the outward normal to the feature on the solar surface and the direction of the line-of-sight of the SDO spacecraft.

We can thus recover the full magnetic field strength by dividing by $\mu_{ij}$:
 \begin{equation}
B_{{\rm r}, ij} = B_{{\rm obs}, ij} / \mu_{ij}.
\end{equation}

The noise level in HMI magnetograms is a function of $\mu$ \citep{2013A&A...550A..95Y}. It is lowest for pixels in the centre of the CCD, where it is close to 5\,G, and increases towards the edges and reaches 8\,G at the solar limb. For our analysis we assume that the noise level is constant throughout the image with a conservative value $\sigma_{B_{{\rm obs}, ij}}$ = 8\,G, in agreement with the results of \citet{2013A&A...550A..95Y}.
We therefore set $B_{{\rm obs}, ij}$ and $B_{{\rm r}, ij}$ to 0 for all pixels with a line-of-sight magnetic field measurement ($B_{{\rm obs}, ij}$) below this value.

\subsection{Identifying quiet-Sun regions, faculae \& sunspots}\label{explain}

As is routinely done in solar work, we do not consider pixels that are very close to the limb ($\mu_{ij}$ < 0.1), as the limb darkening model becomes unreliable on the very edge of the Sun. This affects about 1\% of the pixels of the solar disc.

The first three panels of Figure~\ref{pictures} show an SDO/HMI flattened intensitygram, line-of-sight Dopplergram and unsigned radial magnetogram for a set of images taken on 2011, November 10, after removing the contributions from spacecraft motion and solar rotation. We identify quiet-Sun regions, faculae and sunspots by applying magnetic and intensity thresholds.

\begin{description}
\item[\emph{- Magnetic threshold:}] 
The distribution of pixel unsigned observed magnetic field strength as a function of pixel flattened intensity is shown in Figure~\ref{patchwork}. In the top histogram and main panel, we see that the distribution of magnetic field strength falls off sharply with increasing field strength. The vast majority of pixels are clustered close to the noise level: these pixels are part of the quiet-Sun surface. Note that the fragmented distribution of pixels with magnetic field strength less than a few G arises from the numerical precision in the SDO/HMI images. This is not an issue in our analysis, however, as these pixels are well below the magnetic noise threshold and we set their field value to zero.
We separate active regions from quiet-Sun regions by applying a threshold in unsigned radial magnetic field strength for each pixel. \citet{2013A&A...550A..95Y} investigated the intensity contrast between the active and quiet photosphere using SDO/HMI data, and found an appropriate cutoff at:
\begin{equation}
|B_{{\rm r}, ij}| > 3\, \sigma_{B_{{\rm obs}, ij}} / \mu_{ij},
\end{equation}
where $\sigma_{B_{{\rm obs}, ij}}$ represents the magnetic noise level in each pixel (see last paragraph of Section~\ref{magnetic}). 
As in \citet{2013A&A...550A..95Y}, we exclude \emph{isolated} pixels that are above this threshold as they are likely to be false positives.
We can thus write:
\begin{equation}
|B_{{\rm r, thresh}, ij}| = 24\, {\rm G} \,/\mu_{ij}.
\end{equation}
The combined effects of imperfect limb-darkening correction and the $\mu$-correction for the radial magnetic field strength at the very edge of the solar disc ($\mu_{ij}$ < 0.3) result in a rim of dark pixels being identified as sunspot pixels. In order to avoid this, we set all such pixels as quiet-Sun elements. In any case, sunspots become invisible near the edge of the solar disc because of the Wilson depression \citep{1958AuJPh..11..177L}, so our cut should not affect the identification of real sunspot pixels.

\begin{figure}
\centering
\includegraphics[width = 0.5\textwidth]{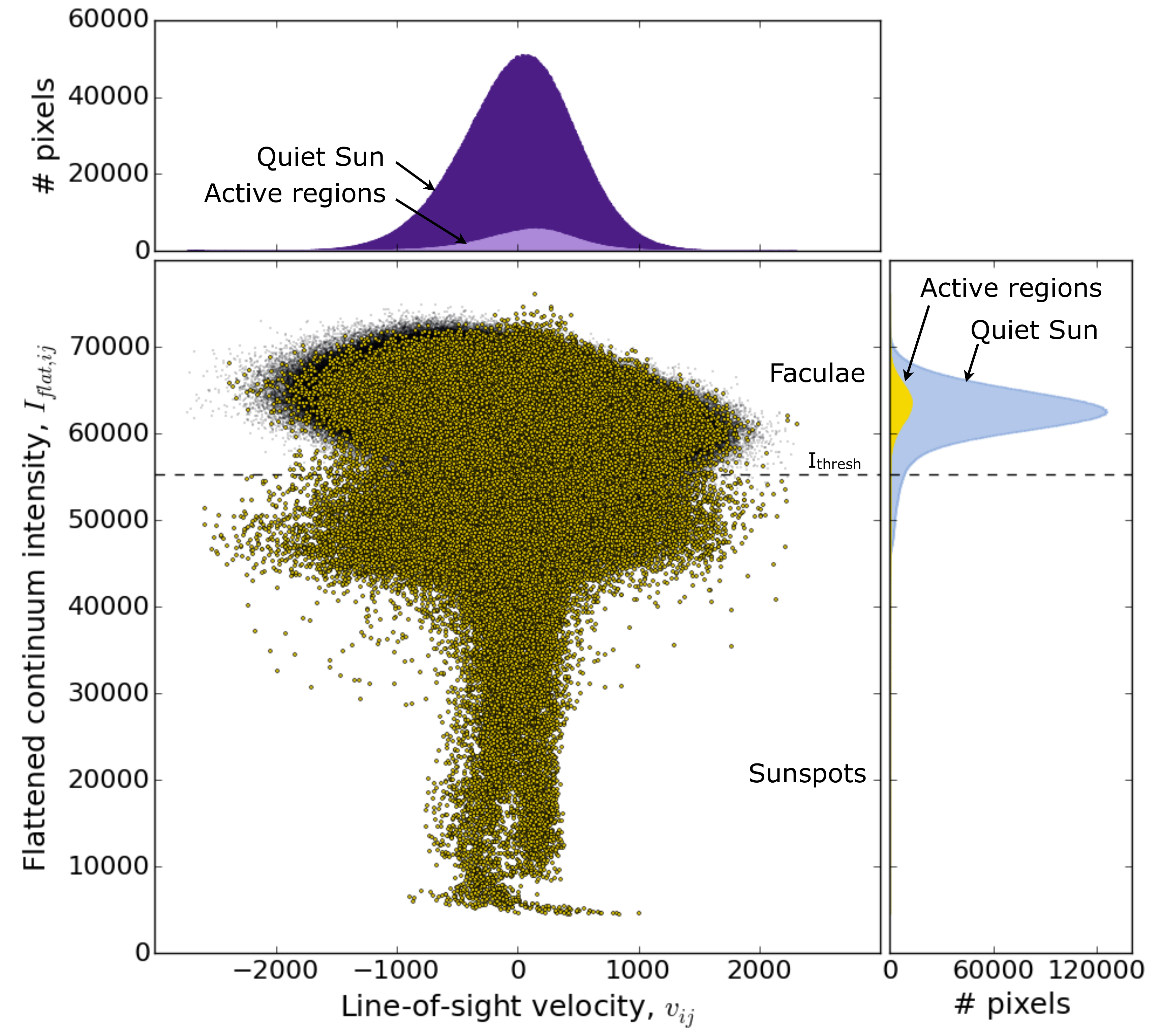}
\centering
\caption{Pixel line-of-sight velocity, $v_{ij}$ (in~m~s$^{-1}$) , as a function of flattened intensity $I_{{\rm flat}, ij}$, for the Sun on 2011, November 10 at 00:01:30 UTC. The top and right histograms show the distributions of $v_{ij}$ and $I_{{\rm flat}, ij}$, respectively, in bins of 1000. In the case of active pixels (yellow dots), the line-of-sight velocity is invariant with pixel brightness. For quiet-Sun pixels (black dots), however, brighter pixels are blueshifted while fainter pixels are redshifted: this effect arises from granular motions.}
\label{jellyfish}
\end{figure}

\item[\emph{- Intensity threshold:}] 
The distribution of line-of-sight velocity as a function of pixel flattened intensity is shown in Figures~\ref{patchwork} and~\ref{jellyfish}.
The main panel allows us to further categorise active-region pixels into faculae and sunspots (umbra and penumbra). We apply the intensity threshold between faculae and spots of \citet{2013A&A...550A..95Y}:
\begin{equation}
I_{\rm thresh}\,=\,0.89\,\hat{I}_{\rm quiet},
\end{equation}
where $\hat{I}_{\rm quiet}$ is the mean pixel flattened intensity over quiet-Sun regions. It can be calculated by summing the flattened intensity of each pixel:
\begin{equation}
\hat{I}_{{\rm quiet}} = \frac{\sum_{ij} I_{{\rm flat}, ij} \, W_{ij}}{\sum_{ij} W_{ij}},
\end{equation}
where the weighting factor $W_{ij}$ is set to 1 if $|B_{{\rm r}, ij}| < |B_{{\rm r, thresh}, ij}|$, 0 otherwise.

\end{description}
In the main panel of Figure~\ref{jellyfish}, quiet-Sun pixels are plotted in black, while active-region pixels are overplotted in yellow. 
The last panel of Figure~\ref{pictures}, which shows the thresholded image according to these $I_{{\rm flat}, ij}$ and $|B_{{\rm r}, ij}|$ criteria, confirms that these thresholding criteria are effective at identifying sunspot and faculae pixels correctly.

\subsection{Surface velocity flows in sunspot penumbrae and umbrae}
\begin{figure}
\centering
\includegraphics[width = 0.3\textwidth]{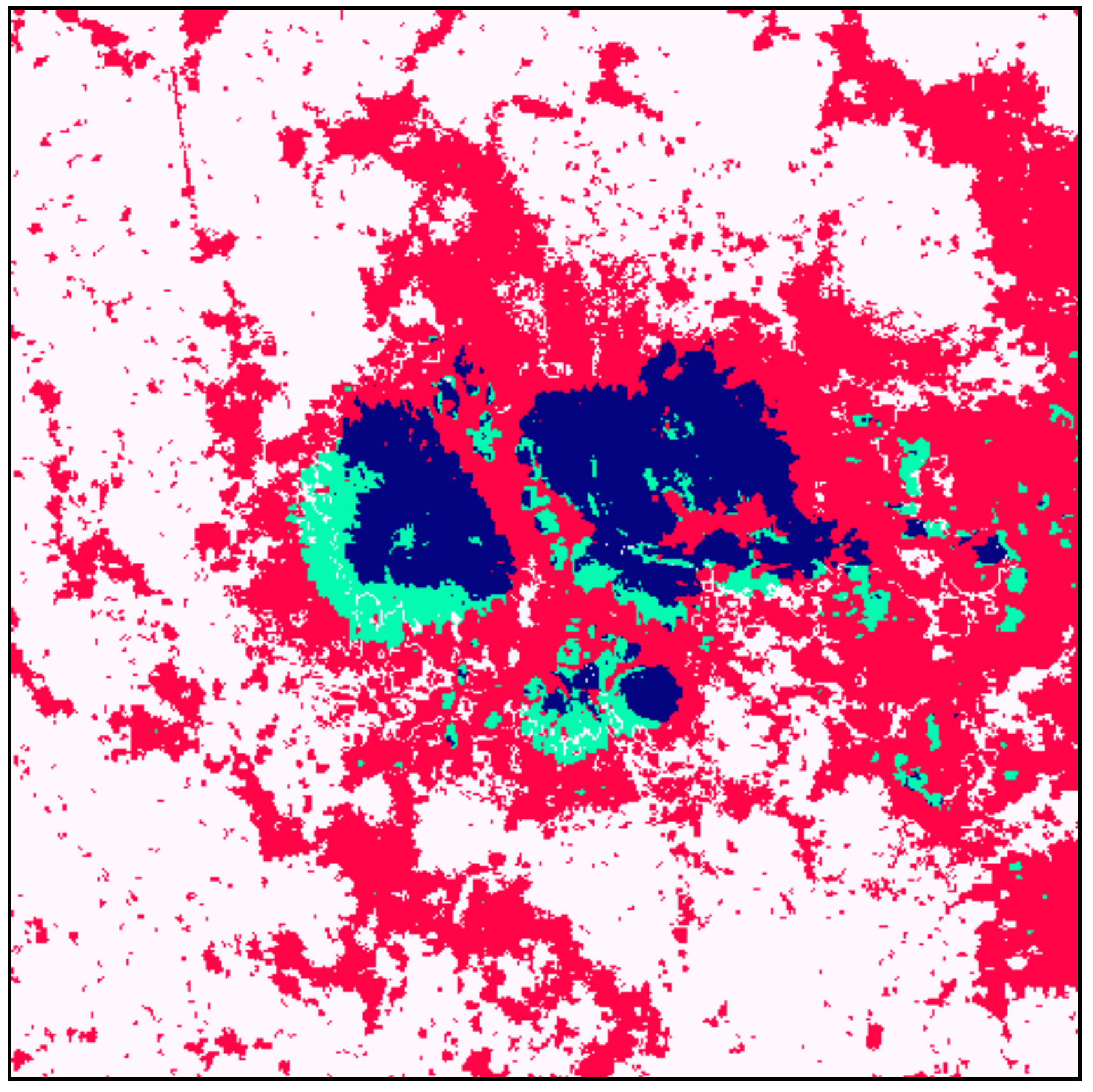}
\centering
\caption{Zoom-in on the largest sunspot group on the approaching hemisphere of the Sun observed on 2011, November 10 at 00:01:30 UTC. Red pixels represent faculae, while the sunspot pixels are colour-coded according to the direction of their velocity flows, thereby revealing the presence of Evershed flows: dark blue pixels are directed towards the observer ($v < 0$), while light cyan are directed away from the observer ($v > 0$).}
\label{sunspot}
\end{figure}

The range of velocities in the penumbral pixels (which form the horizontal oval of yellow points just below $I_{\rm thresh}$ in Figure~\ref{jellyfish}) is large owing to the Evershed effect  \citep{1909MNRAS..69..454E}. 
The darker umbral pixels have a narrower velocity distribution,
allowing the umbral flows that feed the Evershed effect to be resolved into a blueshifted and redshifted velocity component.
The much broader range of velocities seen in Figure~\ref{jellyfish} for penumbral \emph{vs.} umbral pixels confirms that these velocity flows accelerate with distance from the middle of the spot \citep{1909MNRAS..69..454E}.
This effect is highlighted in Figure~\ref{sunspot}, which shows a zoom-in on the largest sunspot group in the images of Figure~\ref{pictures}. The Evershed flows are tangential to the surface. They will be most visible for sunspots located away from disc centre, where a larger proportion of them will be directed along our line-of-sight.

The sunspot group illustrated in Figure~\ref{sunspot} is located near the approaching side of the Sun. The flows in its left half are directed away from the observer, while in its right half they are directed towards the observer.

\subsection{Decomposing the Sun's RV into individual feature contributions}
\begin{figure}
\centering
\includegraphics[width=0.35\textwidth]{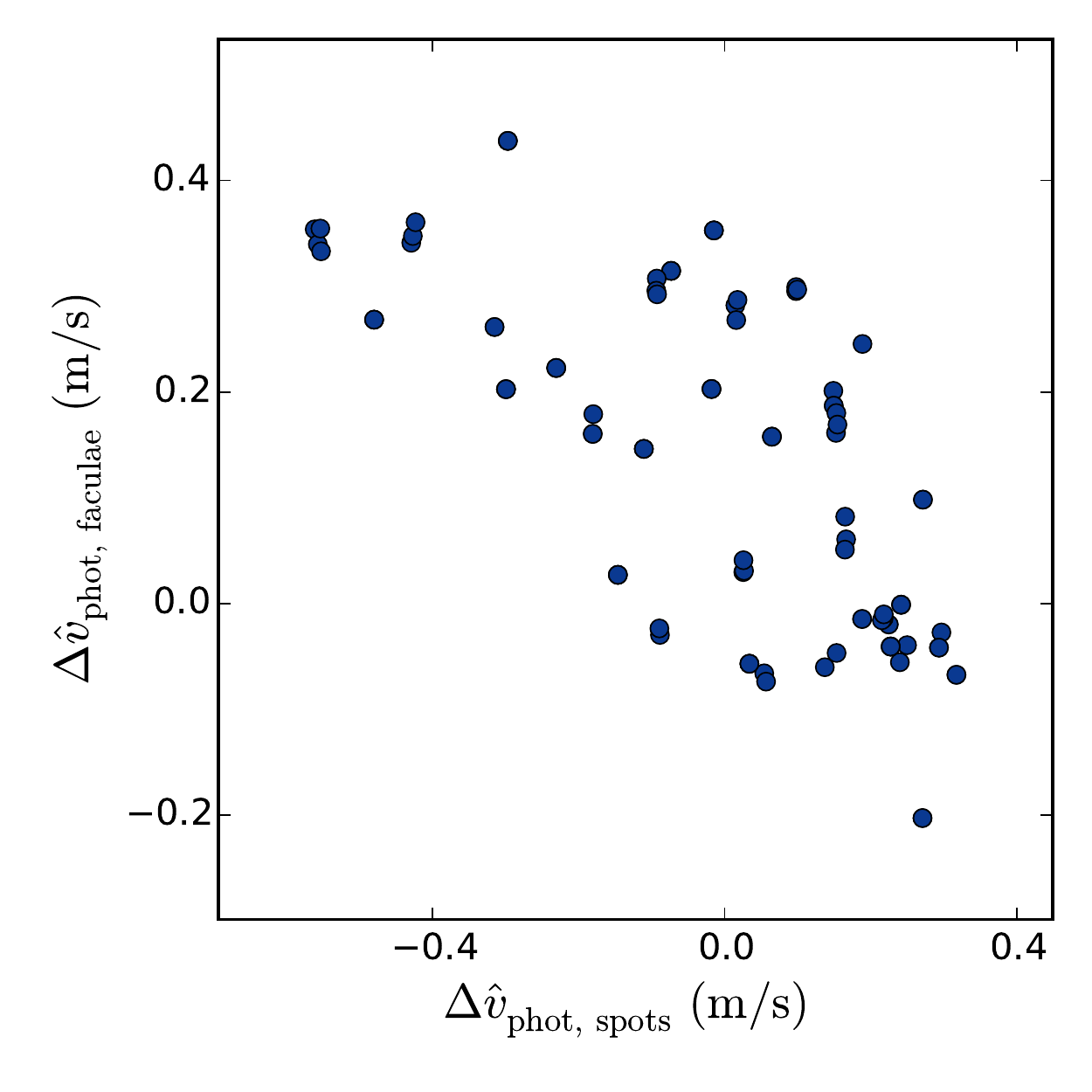}
\centering
\caption{Correlation diagram showing the relationship between the rotational Doppler imbalances resulting from sunspots (horizontal axis) and faculae (vertical axis), if computed separately. The Spearman correlation coefficient is -0.69. The variations are only partially anti-correlated, reflecting both the tendency of faculae to be spatially associated with sunspot groups and the existence of faculae in bright networks not associated with sunspots. The values of these two basis functions are given in the Supplementary Files that are available online.}
\label{faculae}
\end{figure}

The total disc-integrated RV of the Sun is the sum of all contributions from the quiet-Sun, sunspot and faculae/plage regions.
The "quiet" parts of the Sun's surface are in constant motion due to granulation, while sunspots and faculae induce RV variations via two processes (\emph{cf.} introduction):
\begin{description}
\item [\emph{- Photometric effect:}] 
as the Sun rotates, the presence of dark spots or bright faculae on the solar surface breaks the Doppler balance between the approaching (blueshifted) and receding (redshifted) hemispheres. 
We calculated the contribution of dark spots and bright faculae separately and our findings confirm those of  \citet{Meunier:2010hc}, who found that this residual signal is approximately equal to the photometric contribution from sunspots. 
We note, however, that this is merely a coincidence arising from the specific geometrical configuration and ratio of sunpots to faculae/plage on the Sun. In other words, this assumption may not be valid for other stars with different spot to faculae configurations and/or filling ratios. We therefore decide to include the effect of faculae in this study (even though it is effectively negligible in the case of the Sun).
Because the two contributions are correlated (see Figure~\ref{faculae}), we must account for them in a single term $\Delta \hat{v}_{\rm phot}$, which we describe in Section~\ref{rvr}.

\item [\emph{- Convective effect:}] 
sunspots and faculae are strongly magnetised features that inhibit convective motions. Sunspots, which cover a small area of the solar surface and contribute little flux, have a very small contribution. In the case of faculae, however, this contribution is large and is expected to be the dominant contribution to the total solar RV variations \citep{Meunier:2010hc, 2010A&A...519A..66M}. We compute this contribution $\Delta \hat{v}_{\rm conv}$ in Section~\ref{rvc}.
\end{description}

We can thus write the total disc-integrated RV of the Sun as:
\begin{equation}
\hat{v} = \hat{v}_{\rm quiet} + \Delta \hat{v}_{\rm phot} + \Delta \hat{v}_{\rm conv},
\end{equation}
where $\hat{v}_{\rm quiet}$ represents the velocity contribution of convective motions in quiet-Sun pixels, which we calculate in Section~\ref{vquiet}. 

\subsubsection{Velocity contribution of convective motions in quiet-Sun regions}\label{vquiet}
We estimate the average RV of the quiet Sun by summing the intensity-weighted velocity of non-magnetised pixels, after removing the spacecraft motion and the Sun's rotation:
\begin{equation}
\hat{v}_{{\rm quiet}} = \frac{\sum_{ij} (v_{ij} - \delta v_{{\rm sc}, ij} - \delta v_{{\rm rot}, ij}) \, I_{ij} \, W_{ij}}{\sum_{ij} I_{ij} \, W_{ij}}.
\end{equation}
For this calculation, the weights are defined as in:
\begin{equation}
\begin {aligned}
W_{ij} = 1 \, \, {\rm if}\, \, |B_{{\rm r}, ij}| < |B_{{\rm r, thresh}, ij}|, \\
W_{ij} = 0 \, \, {\rm if}\, \, |B_{{\rm r}, ij}| > |B_{{\rm r, thresh}, ij}|. \\
 \end{aligned}
\end{equation}

This velocity field is thus averaged over the vertical motions of convection granules on the solar surface. Hot and bright granules rise up to the surface, while cooler and darker fluid sinks back towards the Sun's interior. This process is visible in the main panel of Figure~\ref{jellyfish}: quiet-Sun pixels (black dots) are clustered in a tilted ellipse.
The area of the upflowing granules is larger than that enclosed in the intergranular lanes, and the granules are carrying hotter and thus brighter fluid. This results in a net blueshift, as seen in Figure~\ref{jellyfish}.



\subsubsection{Rotational Doppler imbalance due to dark sunspots and bright faculae} \label{rvr}
This velocity perturbation can be obtained by summing the line-of-sight velocity of sunspot pixels corrected for the spacecraft's motion, and weighted by the flux deficit or excess produced by a dark spot or a bright plage, respectively:
\begin{equation}
\Delta \hat{v}_{{\rm phot}} = \frac{\sum_{ij} (v_{ij} - \delta v_{{\rm sc}, ij}) \, (I_{ij} - \hat{K} \, L_{ij}) \, W_{ij}}{\sum_{ij} I_{ij}}.
\end{equation}
The weights $W_{ij}$ are set to 1 for pixels with $|B_{{\rm r}, ij}| > |B_{{\rm r, thresh}, ij}|$ (otherwise they are set to 0).

As illustrated in Figure~\ref{faculae}, the photometric effects of sunspots and faculae are quite similar in amplitude and roughly anti-correlated, due to their opposite flux signs. However, because they are not strictly anti-correlated they sum into a net signal that has an amplitude similar to the photometric effect of sunspots \citep{Meunier:2010hc}.
The value of $\Delta \hat{v}_{{\rm phot}}$ at each time of the HARPS observations is listed in Table~A1. From the SDO images we derive an rms amplitude of 0.17~m~s$^{-1}$ for this signal, which is smaller than the observational uncertainties of the HARPS velocities. This value is slightly lower than that found by \citet{Meunier:2010hc} during the peak of the Sun's activity cycle, of 0.42~m~s$^{-1}$, but remains broadly consistent with their results given the small amplitude of this signal.

\subsubsection{Suppression of convective blueshift from active regions}\label{rvc}
The suppression of granular blueshift induced by magnetically active regions ($|B_{{\rm r}, ij}| > |B_{{\rm r, thresh}, ij}|$, predominantly faculae) is:
\begin{equation}
\Delta \hat{v}_{{\rm conv}} = \hat{v} - \hat{v}_{{\rm quiet}} - \hat{v}_{{\rm phot}}.
\end{equation}

We measure the total disc-averaged velocity of the Sun $\hat{v}$ by summing the velocity contribution of each pixel $ij$, weighted by their intensity $I_{ij}$, after subtracting the spacecraft motion and solar rotation:
\begin{equation}
\hat{v} = \frac{\sum_{ij} (v_{ij} - \delta v_{{\rm sc}, ij} - \delta v_{{\rm rot}, ij}) \, I_{ij}}{\sum_{ij} I_{ij}}.
\end{equation}

The value of $\Delta \hat{v}_{{\rm conv}}$ at each time of the HARPS observations is listed in Table~A1.
The rms amplitude of this basis function (unscaled) is 1.30~m~s$^{-1}$, which is consistent with the value of 1.39~m~s$^{-1}$ computed by \citet{Meunier:2010hc} during the peak of the solar actvity cycle.

\section{Reproducing the RV variations of the Sun}\label{scaling}\label{reproduce}

\subsection{Total RV model}\label{model}
We combine our model of Vesta's rotational RV signal (presented in Section~\ref{axial}) with the two magnetic activity basis functions determined in Sections~\ref{rvc} and~\ref{rvr}, in order to reproduce the RV variations seen in the HARPS observations. 
The final model has the form: 
\begin{equation}
\begin {aligned}
\Delta RV_{\rm model}(t) = A\,\Delta \hat{v}_{{\rm phot}}(t) + B\,\Delta \hat{v}_{{\rm conv}}(t) + \Delta RV_{\rm vesta}(t) 
+ RV_0. 
 \end{aligned}
\end{equation}

\begin{figure}
\centering
\includegraphics[width=0.48\textwidth]{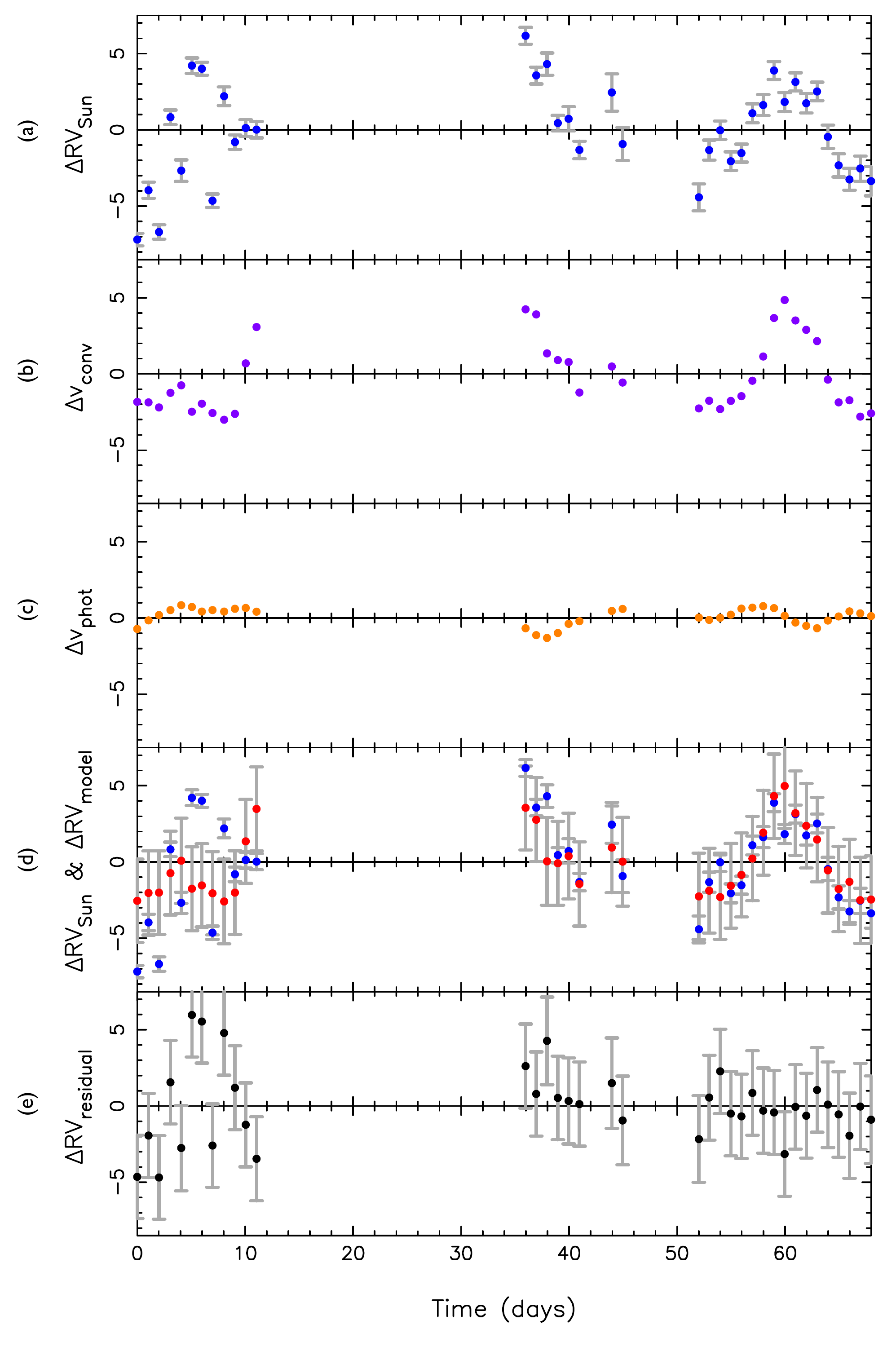}
\centering
\caption{\emph{Panel (a):} HARPS RV variations of the Sun as-a-star, $\Delta RV_{\mathrm{Sun}}$; \emph{Panel (b):} Scaled basis function for the suppression of convective blueshift, $\Delta \hat{v}_{\mathrm{conv}}$, derived from SDO/HMI images; \emph{Panel (c):} Scaled basis function for the rotational Doppler imbalance due to spots and faculae, $\Delta \hat{v}_{\mathrm{phot}}$; \emph{Panel (d):} total RV model, $\Delta RV_{\mathrm{model}}$ (red/lighter shade, with errors including additional variance $s$), overlaid on top of the HARPS RV variations (blue/darker shade points); \emph{Panel (e):} residuals obtained after subtracting the model from the observations. All RVs are in m~s$^{-1}$. Note that the scale of the $y$-axis is different to that used in Figure~\ref{two}. The values of these nightly-binned timeseries are provided in the Supplementary Files available online.}
\label{all}
\end{figure}

We carry out an optimal scaling procedure in order to determine the scaling factors ($A, B, C$ and $S$) of each of the contributions, as well as the constant offset $RV_0$ and a constant variance term $s^2$ added in quadrature to the observational errors.
This variance term will account for any remaining uncorrelated noise arising from granulation motions. The SDO/HMI images were exposed for 45 seconds each, while the HARPS observations were exposed for 15 minutes and binned nightly. We expect these differences to result in a small night-to-night uncorrelated noise contribution. In addition, this variance term will absorb any residual signal of Vesta's axial rotation, which will naturally not be completely sinusoidal.

For each value of $s$, the basis functions are orthogonalised by subtracting their inverse-variance weighted means prior to performing the scaling.
We determine the maximum likelihood via a procedure similar to the one described in \citet{CollierCameron:2006fx}, whereby we maximise the likelihood $\mathcal{L}$ of the solution:
\begin{equation}
\ln \mathcal{L}  = - \frac{n}{2} \ln(2\pi) -\frac{1}{2} \, \chi^2 - \frac{1}{2} \, \sum_{i=1}^{i=n} \ln(\sigma^2_i+s^2)
\end{equation}
with respect to $s$, where $\chi^2$ is the chi-squared value of the $n$ HARPS data points, with uncertainties $\sigma_i$.
This procedure is applied to the unbinned (not nightly-averaged) HARPS dataset, in order to determine the appropriate scaling coefficients ($C$ and $S$) for Vesta's axial rotation (see Equation~\ref{eqn6}). The total amplitude of the modulation induced by Vesta's rotation is equal to 2.39~m~s$^{-1}$, which is of the same order as the amplitude we estimated in Section~\ref{axial}.

\subsection{Agreement between HARPS observations and SDO-derived model}
After all the scaling coefficients were determined, we grouped the observations in each night by computing the inverse variance-weighted average for each night. The final model is shown in Figure~\ref{all}.

We list the best-fit values of the scaling parameters for each of the basis functions in Table~\ref{table}. We note that the values of $A$ and $B$, which represent the difference in response of HARPS and SDO to the Doppler imbalance and the suppression of convective blueshift, differ from unity.
The HMI/SDO images are based on measurements of a single spectral line, namely the Fe {\sc I} 6173\AA~line, which may not be representative of the several thousand lines from which the HARPS RVs are derived. 
The value of $\Delta \hat{v}_{\mathrm{phot}}$ is determined by the contrast between the magnetic elements (spots, faculae) and the quiet-Sun photosphere. Most of the lines in the HARPS mask lie blueward of Fe {\sc I} 6173\AA\ where the contrast is greater for both types of feature. We therefore expect to have to scale the SDO-derived $\Delta \hat{v}_{\mathrm{phot}}$ by a factor of order, but greater than, unity. We measure it to be $2.45 \pm 2.02$; the amplitude of $\Delta \hat{v}_{\mathrm{phot}}$ is so small that $A$ cannot be distinguished from unity when the additional variance $s$ is taken into account.
The strength of the convection inhibition changes with line depth (\emph{cf.} \citet{2009ApJ...697.1032G} and \citet{2010A&A...519A..66M}). We find that the HARPS response to suppression of granular blueshift is $1.85 \pm 0.27$ times greater than we predict from SDO.

\begin{figure}
\centering
\includegraphics[width = 0.35\textwidth]{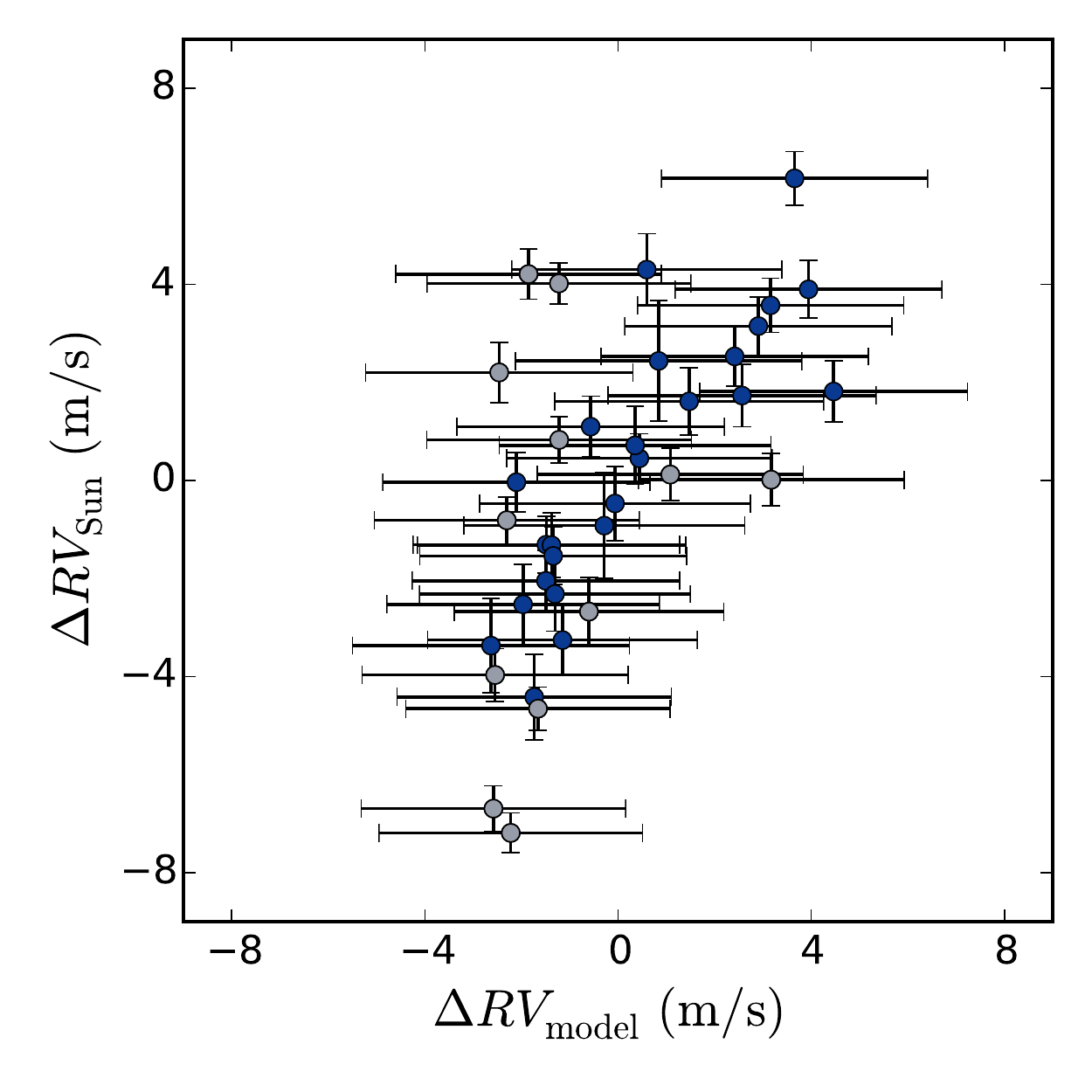}
\centering
\caption{HARPS RV variations of the Sun as-a-star \emph{vs.} our model derived from SDO/HMI images. Observations from the first part of the run are highlighted in a lighter shade. }
\label{comparison}
\end{figure}

 We show a plot of the observed HARPS RVs ($\Delta RV_{\mathrm{Sun}}$) \emph{vs.} our SDO-modelled RVs ($\Delta RV_{\mathrm{model}}$) in Figure~\ref{comparison}, in a similar fashion to Figure 9 of \citet{2010A&A...519A..66M}. We see good agreement between the data and the model. The Spearman correlation coefficient is 0.64 for the full dataset and 0.87 when considering only the second part of the run. This is close to, although not as good as the correlation coefficient of 0.94 found by \citet{2010A&A...519A..66M} between simulated velocities of \cite{Meunier:2010hc} and MDI/SOHO Dopplergrams.

\begin{table}
\centering
\caption{Best-fit parameters and rms amplitudes resulting from the optimal scaling procedure.}
 \begin{tabular}{@{}llll@{}}
 Parameter & Value & Basis function & rms amplitude \\
 &&&(unscaled) \\
 \hline
 \vspace{0.01 cm}\\
$A$ & $2.45 \pm 2.02 $ 			& $\Delta \hat{v}_{{\rm phot}}$ 	& 0.17~m~s$^{-1}$	  \\
$B$ & $ 1.85 \pm 0.27 $			& $\Delta \hat{v}_{{\rm conv}}$	& 1.30~m~s$^{-1}$	 \\	
$C$ & $ 2.19 \pm 0.42 $ 			& $\cos(2\pi - \lambda)$	& 0.74~m~s$^{-1}$	\\
$S$ & $ 0.55 \pm 0.47 $			& $\sin(2\pi - \lambda)$	& 0.66~m~s$^{-1}$	\\
\vspace{0.01 cm}\\
$RV_0$ & $ 99.80 \pm 0.28$~m~s$^{-1}$\\
$s$ & $2.70$~m~s$^{-1}$ \\
\vspace{0.01 cm}\\
\hline
\label{table}
\end{tabular}
\end{table}


Panel (e) of Figure~\ref{all} shows the residuals remaining after subtracting
the total model $\Delta RV_{\mathrm{model}}$ from the HARPS observations of the Sun as-a-star $\Delta RV_{\mathrm{Sun}}$. We see that the residuals are within the level of the error bars, when considering an extra variance term $s$ of 2.7~m~s$^{-1}$.
Maximum-likelihood analysis of the additional uncorrelated noise during the first 
part of the run (nights 0-11) yields an additional variance with 
rms amplitude 4.0~m~s$^{-1}$, while the second part (nights
36-68) has an additional noise rms of 1.5~m~s$^{-1}$. As discussed in Section~\ref{instrument}, we attribute the excess scatter in the first block of nights to airmass-dependent guiding errors, which would have been more important in the first part of the run, where the observations within each night were taken at very different airmasses.
The additional noise rms of 1.5~m~s$^{-1}$ in the second part of the run is consistent with the rms due to solar granulation and super-granulation, ranging between 0.28 and 1.12~m~s$^{-1}$, as recently found by \citet{2015A&A...583A.118M}.

\begin{figure*}
\centering
\includegraphics[width = 0.8\textwidth]{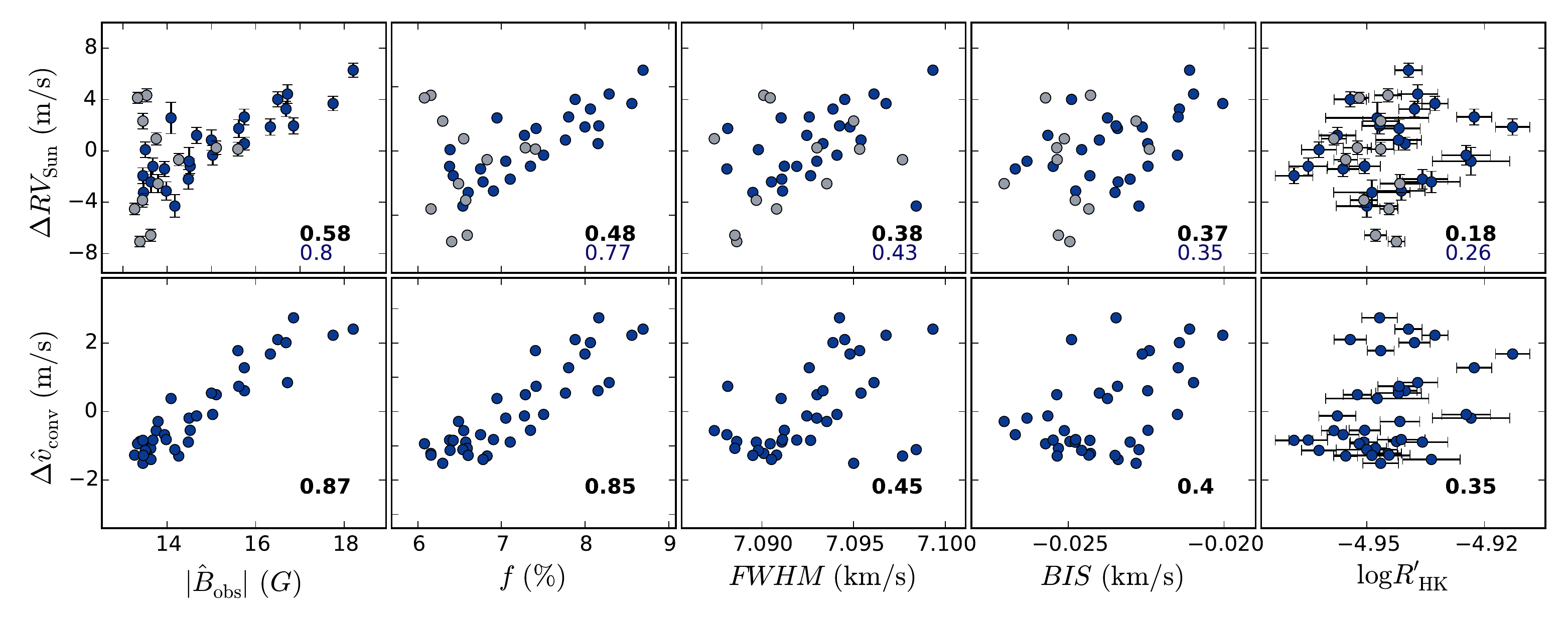}
\centering
\caption{Correlation plots of the nightly-averaged HARPS RV variations of the Sun as-a-star $\Delta RV_{{\rm Sun}}$ and suppression of convective blueshift $\Delta \hat{v}_{{\rm conv}}$ against \emph{(from left to right)}: the disc-averaged observed magnetic flux $|\hat{B}_{\rm obs}|$ (G), filling factor $f$ (\%), FWHM (km~s$^{-1}$), BIS (m~s$^{-1}$) and $\log(R'_{\rm HK})$. Observations from the first part of the run are highlighted in a lighter shade. 
Spearman correlation coefficients are displayed in the bottom-right corner of each panel: for the full observing run (in bold and black), and for the second part of the run only (in blue). }
\label{correlation}
\end{figure*}

\subsection{Relative importance of suppression of convective blueshift and rotational velocity imbalance}
We see that the activity-induced RV variations of the Sun are well reproduced by a scaled sum of the two basis functions, $\hat{v}_{{\rm conv}}$ and $\hat{v}_{{\rm phot}}$ (shown in Figure~\ref{all} panels (b) and (c), respectively).
As previously predicted by \cite{Meunier:2010hc}, we find that the suppression of convective blueshift plays a dominant role (rms of 2.40~m~s$^{-1}$). This was also found to be the case for CoRoT-7, a main sequence G9 star with a rotation period comparable to that of the Sun \citep{Haywood:2014hsa}. The relatively low amplitude of the modulation induced by sunspot flux-blocking (rms of 0.17~m~s$^{-1}$) is expected in slowly-rotating stars with a low $v \sin i$ \citep{Desort:2007dt}.

\subsection{Zero point of HARPS}\label{offset}
The wavelength adjustments that were applied to the HARPS RVs were based on precise prior dynamical knowledge of the rate of change of distance between the Earth and Vesta, and between Vesta and the Sun. The offset $RV_0 = 99.80 \pm 0.28$ m~s$^{-1}$ thus represents the zero point of the HARPS instrument, including the mean granulation blueshift for the Sun. 
A previous study of integrated sunlight reflected by the Moon, by \citet{Molaro:2013je}, determined a value of $102.2 \pm 0.86$~m~s$^{-1}$, which is significantly larger than our value. They did not account for the effect of sunspots and faculae, however, so their result could be affected by activity-induced solar variations at that time.

\section{Towards better proxies for RV observations}\label{proxy}

\subsection{Spatial distributions of sunpots and faculae}
\citet{Aigrain:2012} have shown that it is possible to predict the rotational Doppler imbalance due to photospheric surface brightness inhomogeneities from a simultaneous high-precision optical lightcurve. If one further assumes that faculae/plage regions are co-spatial with spot groups, then they can also predict the form of the RV variation caused by suppression of granular blueshift. 
A recent analysis of the active host star CoRoT-7 by \citet{Haywood:2014hsa} modelled activity-induced RV variations via the \emph{FF'} method of \citet{Aigrain:2012}. The predicted Doppler imbalance was much smaller than the observed activity-driven RV variations. The associated suppression of convective blueshift was of larger amplitude than, and partially correlated with, the observed RVs. The residuals, however, had a similar amplitude and shared the covariance properties of the star's (simultaneous) lightcurve.

The present study provides a natural explanation of this mismatch: on the Sun, the faculae are not perfectly co-spatial with sunspot groups. Indeed, Figure~\ref{faculae} shows us that the location of sunspot groups give an incomplete prediction of the facular coverage. 
Since the suppression of granular blueshift is the dominant process at play in slowly-rotating stars such as CoRoT-7 and the Sun, it is therefore important to develop proxies that are directly sensitive to the distribution of faculae on the stellar surface. 

\subsection{Correlations between RV and traditional activity indicators}
Figure~\ref{correlation} presents the correlations between $\Delta RV_{{\rm Sun}}$, $\Delta \hat{v}_{{\rm conv}}$ and the following activity indicators: the full-disc magnetic flux $|\hat{B}_{\rm obs}|$ and filling factor $f$ computed from the SDO/HMI images, and the observed FWHM, BIS, and $\log(R'_{\rm HK})$ derived from the HARPS DRS reduction pipeline. 
We computed the Spearman correlation coefficent to get a measure of the degree of monotone correlation between each variable (the correlation between two variables is not necessarily linear, for example between RV and BIS). The coefficients are displayed in each panel of Figure~\ref{correlation}, both including and excluding the observations made in the first part of the run, which show a lot of intra-night scatter.

We do not show similar correlation plots for $\Delta \hat{v}_{{\rm phot}}$ because we do not find any significant correlations with any of the activity indicators; this is expected since $\Delta \hat{v}_{{\rm phot}}$ is such that it crosses zero when the surface covered by spots and/or faculae is at a maximum, \emph{i.e.} when they are in the middle of the stellar disc ($\Delta \hat{v}_{{\rm phot}}$ is of course still related to $|\hat{B}_{\rm obs}|$, the FWHM, BIS, and $\log(R'_{\rm HK})$, but the Spearman coefficient is close to zero).

We note that the relatively weak correlation between the observed RVs and the BIS is not completely unexpected in the case of the Sun, which has a $v \sin i$ of about 2~km~s$^{-1}$. The line profile distortions induced by solar activity, including those measured by the BIS, will therefore be smaller than the resolution of HARPS that is close to 2.5~km~s$^{-1}$. In other words, the resolution of HARPS is not adequate to fully resolve the BIS variations in a star rotating as slowly as the Sun, which could reduce the correlation coefficients computed with the BIS. 

 \begin{figure}
\centering
\includegraphics[width = 0.485\textwidth]{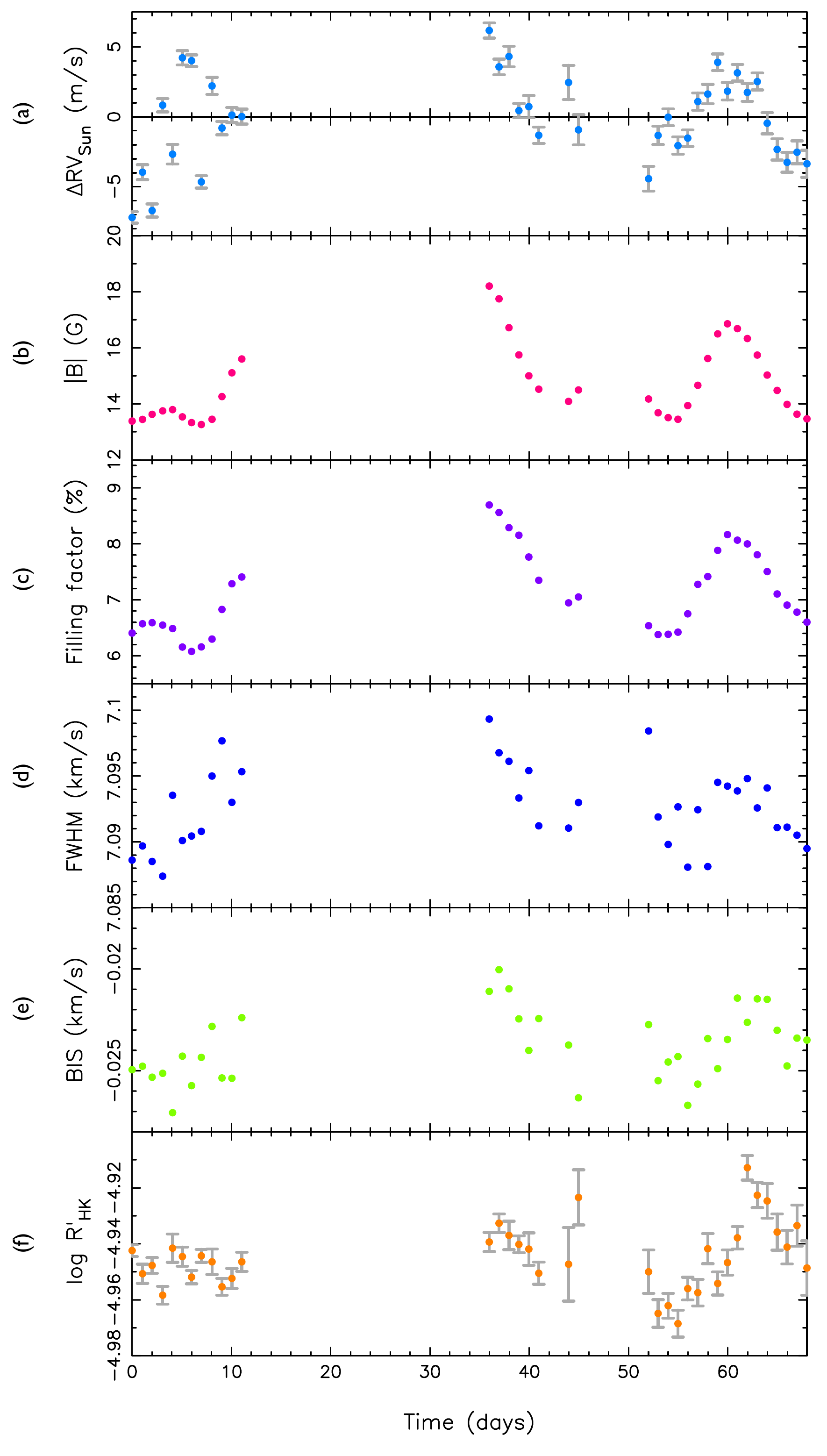}
\centering
\caption{\emph{Panel (a):} HARPS RV variations of the Sun as-a-star; \emph{panels (b) and (c):} time series of the disc-averaged line-of-sight magnetic flux $|\hat{B}_{\rm obs}|$ and filling factor $f$, respectively, determined from the SDO/HMI magnetograms; \emph{panels (d), (e) and (f):} time series of the FWHM, BIS and $\log(R'_{\rm HK})$, respectively, determined from the HARPS DRS reduction pipeline. The values of these nightly-binned timeseries are provided in the Supplementary Files available online.}
\label{variations}
\end{figure}

We find a relatively weak correlation between the observed RVs and the chromospheric flux index $\log(R'_{\rm HK})$, with a Spearman correlation coefficient of 0.26 for the second half of the run (0.18 for the full run).

Although naively one might expect $\log(R'_{\rm HK})$ to be a good predictor of plage filling factor, and hence of the convective RV component, there are several physical factors that might reasonably be expected to degrade the correlation over short time scales (our dataset spans 2-3 solar rotations). Foreshortening and limb darkening affect the Ca {\sc II} HK emission cores and the brightness-weighted line-of-sight granular velocities in different ways. Near the limb, the Ca {\sc II} emission in plages originates in higher, hotter regions of the chromosphere and remains bright. The limb darkening of facular pixels is less than that of quiet-sun pixels, but near the limb the line-of-sight component of the radial motion of bright granule cores is reduced by foreshortening. The disc-averaged line-of-sight magnetic field strength is attenuated by both foreshortening and limb darkening in approximately the same way as the radial motion of the granular flow. This may explain why, even though the RVs and $\log(R'_{\rm HK})$ values were measured simultaneously from the same spectra, the correlation between the variations in $\log(R'_{\rm HK})$ and suppression of granular blueshift appears weak. 
In their study of long-term solar RV variations spanning over 8 years, \citet{Lanza2015} 
find a stronger correlation between $\log(R'_{\rm HK})$ and RV variations, with a Spearman coefficient of 0.357. This positive correlation is in agreement with previous studies of quiet late-type stars \citep{2012A&A...541A...9G,Lovis:2011vq}, and shows that the $\log(R'_{\rm HK})$ may be a more useful proxy for long-term RV variations induced the stellar magnetic cycle.

We note that in Figure~\ref{correlation}, the variations in $\log(R'_{\rm HK})$ look similar to the variations in RV, except that they are shifted by a few days (this is especially noticeable towards the end of the run). The reality and origin of this shift will be the subject of future studies, thanks to the wealth of Sun as-a-star RV observations that are currently being gathered by the solar telescope at HARPS-N.

\subsection{Disc-averaged observed magnetic flux $|\hat{B}_{\rm obs}|$}
We compute the full-disc line-of-sight magnetic flux of the Sun, by summing the intensity-weighted line-of-sight unsigned magnetic flux in each pixel:
\begin{equation}
|\hat{B}_{\rm obs}| = \frac{\sum_{ij} |B_{{\rm obs}, ij}| \, I_{ij}}{\sum_{ij} I_{ij}}.
\end{equation}

The values at the time of each HARPS observation are listed in Table~A1. 
The variations in $|\hat{B}_{\rm obs}|$ are shown in panel (b) of Figure~\ref{variations}, together with the nightly-averaged HARPS RV variations of the Sun as-a-star, in panel (a). We see that the variations in the disc-averaged magnetic flux are in phase with the RV variations, despite the scatter in RV in the first part of the run (discussed in Section~\ref{instrument}). If we only consider the observations in the second part of the run, the Spearman correlation coefficient between $\Delta RV_{{\rm Sun}}$ and $|\hat{B}_{\rm obs}|$ is 0.80 (see Figure~\ref{correlation}).
The correlation is stronger between $\hat{v}_{{\rm conv}}$ and $|\hat{B}_{\rm obs}|$, with a correlation coefficient of 0.87. This is expected since magnetised areas are known to suppress convective blueshift (see \citet{Meunier:2010hc, 2010A&A...519A..66M}). 

 We note that these observations were taken close to the solar cycle maximum, during which the solar photosphere was mostly dominated by a few large sunspot groups (surrounded by facular networks).  \cite{2010A&A...519A..66M} found that the convective shift attenuation is greater in larger structures, since they contain a stronger magnetic field. The relationship between convective shift attenuation and magnetic field, however, is not linear. We thus expect larger RV variations for a few large active regions than when the Sun is in a phase of lower magnetic activity, when the photosphere might be dominated by several smaller structures, even though they would still give the same total flux.
In fact, \citet{Lanza2015} find a much weaker correlation between the mean total magnetic flux measured by the SOLIS VSM instrument and RV (Spearman correlation coefficient of 0.131) over their 8-year dataset, which spans both active and less active phases of the solar cycle.

When compared against correlations with the traditional spectroscopic activity indicators (the FWHM, BIS and $\log(R'_{\rm HK})$), we see that the disc-averaged magnetic flux $|\hat{B}_{\rm obs}|$ is a much more effective predictor of activity-induced RV variations, over the timescale of a few rotation periods.
The averaged magnetic flux may therefore be a useful proxy for activity-driven RV variations as it should map onto areas of strong magnetic fields, which suppress the Sun's convective blueshift. 
The line-of-sight magnetic flux density and filling factor on the visible hemisphere of a star can be measured from the Zeeman broadening of magnetically-sensitive lines \citep{Robinson:1980dm, Reiners:2013kr}. Their product gives the disc-averaged flux density that we are deriving from the solar images.
We note that such measurements are still very difficult to make for other stars than the Sun, because the Zeeman splitting of magnetically sensitive lines is so small that the technique can only be applied to bright, slowly rotating stars. Fortunately, such stars are also the best targets for planet searches.

\subsection{Magnetic filling factor $f$}
In addition to the disc-averaged observed magnetic flux, we also computed the filling factor of magnetic regions on the solar disc. It is weighted by the foreshortening at the location of each pixel, and counted as a fraction of the total pixel count:
\begin{equation}
f = \frac{1}{N_{\rm pix}} \, \sum_{ij} W_{ij},
\end{equation}
where $N_{\rm pix}$ is the total number of pixels in the solar disc and the weight $W_{ij}$ is set to 1 in magnetically active regions, and 0 in the quiet Sun. The variations of the filling factor are shown in panel (c) of Figure~\ref{variations}. As expected, they follow the disc-averaged magnetic flux closely. The correlation with the predicted $\Delta RV_{{\rm conv}}$ and the observed $\Delta RV_{{\rm Sun}}$ is nonetheless weaker than that found for the brightness-weighted line-of-sight magnetic field $|\hat{B}_{\rm obs}|$, since no correction is made for limb darkening.

\section{Conclusion}
In the present analysis, we decomposed activity-induced radial-velocity (RV) variations into identifiable contributions from sunspots, faculae and granulation, based on Sun as-a-star RV variations deduced from  HARPS spectra of the bright asteroid Vesta and high spatial resolution images in the Fe {\sc I} 6173~\AA~line taken with the Helioseismic and Magnetic Imager (HMI) instrument aboard the Solar Dynamics Observatory (SDO). 
We find that the RV variations induced by solar activity are mainly caused by the suppression of convective blueshift from magnetically active regions, while the flux deficit incurred by the presence of sunspots on the rotating solar disc only plays a minor role. We further compute the disc-averaged line-of-sight magnetic flux and show that although we cannot yet measure it with precision on other stars at present, it is a very good proxy for activity-driven RV variations, much more so than the full width at half-maximum and bisector span of the cross-correlation profile, and the Ca II H\&K activity index. These findings are in agreement with the previous works of \citet{Meunier:2010hc} and \citet{2010A&A...519A..66M}.

In addition to the existing 2011 HARPS observations of sunlight scattered off Vesta, there will soon be a wealth of direct solar RV measurements taken with HARPS-N, which will be regularly fed sunlight through a small 2-inch telescope developed specifically for this purpose. A prototype for this is currently being commissioned at HARPS-N (see \citet{2015ApJ...814L..21D}, \citeauthor{Glenday2015} \citetext{in prep.}).
Gaining a deeper understanding of the physics at the heart of activity-driven RV variability will ultimately enable us to better model and remove this contribution from RV observations, thus revealing the planetary signals.

\section*{Acknowledgments}
We wish to thank the referee for their thoughtful comments, which have greatly helped improve the analysis presented in this paper.
RDH gratefully acknowledges STFC studentship grant number
ST/J500744/1, and a grant from the John Templeton Foundation. The opinions expressed in this publication are those of the authors and do not necessarily reflect the views of the John Templeton Foundation.
ACC and RF acknowledge support from STFC consolidated grants numbers
ST/J001651/1 and ST/M001296/1. JL acknowledges support from NASA Origins of the Solar System grant No. NNX13AH79G and from STFC grant ST/M001296/1. The Solar Dynamics Observatory was launched by NASA on 2011, February 11, as part of the Living With A Star program. This research has made use of NASA's Astrophysics Data System Bibliographic Services.

\bibliographystyle{mnras} 
\bibliography{20151028myreferences}

\appendix
\begin{table*}
\caption{HARPS 2011-2012 data for the asteroid Vesta, processed by the HARPS pipeline with the correct barycentric RV and accounting for the relativistic correction, and SDO-derived quantities resulting from our analysis. \emph{From left to right are given:} Julian date (flux-weighted mid-exposure times of observation), RV, the estimated error $\sigma_{\rm RV}$ on the RV measurement, the full width at half-maximum (FWHM) and the line bisector of (BIS) of the cross-correlation function (as defined in \citet{Queloz:2001be}), the Ca II H \& K activity indicator log($R'_{\rm HK}$) and its error $\sigma_{\log(R'_{\rm HK})}$, the timelag $\Delta t$ between the HARPS observations and the time at which SDO observed the same hemisphere of the Sun, the apparent planetographic longitude of Vesta $\lambda$, the values of the basis functions for the sunspot velocity signal, $ \hat{v}_{{\rm phot}}$ and the suppression of granular blueshift, $ \hat{v}_{{\rm conv}}$, the full-disc magnetic flux $|\hat{B}_{\rm obs}|$ and the filling factor of magnetic regions $f$ both computed from the SDO images. An expanded version of this table, together with all Supplementary Files are available electronically at: http://dx.doi.org/10.17630/bb43e6a3-72e0-464c-9fdd-fbe5d3e56a09.}
\begin{center}

{\scriptsize
\begin{tabular}{cccccccccccccc}\label{A1}
Julian Date & RV & $\sigma_{\rm RV}$ & FWHM & BIS & $\log(R'_{\rm HK}$) & $\sigma_{\log(R'_{\rm HK})}$ & $\Delta t$ & $\lambda$ & $ \hat{v}_{{\rm phot}}$ & $ \hat{v}_{{\rm conv}}$ & $|\hat{B}_{\rm obs}|$ & $f$\\

[MJD$_{\rm mid}$\_UTC] & [km~s$^{-1}$] & [km~s$^{-1}$] & [km~s$^{-1}$] & [km~s$^{-1}$] &    &   & [days] & [deg] & [m~s$^{-1}$] & [m~s$^{-1}$] & [G] & [\%] \\
\hline
\rule{0pt}{0ex} \\

2455834.48296192 & 0.08570 & 0.00043 & 7.08902 & -0.02417 & -4.9411 & 0.0024 & 2.8365 & 143.99 & -0.122949 & 6.612382  & 13.390075 & 6.442092 \\
2455834.61321887 & 0.09936 & 0.00039 & 7.0883  & -0.02559 & -4.9435 & 0.0019 & 2.8451 & 351.75 & -0.122949 & 6.612382  & 13.390075 & 6.442092 \\
2455835.53355275 & 0.09477 & 0.00049 & 7.08903 & -0.02478 & -4.9513 & 0.0028 & 2.8901 & 39.73  & -0.372071 & 6.728378  & 13.461877 & 6.553660 \\
2455835.66225392 & 0.09936 & 0.00060 & 7.09069 & -0.02477 & -4.9498 & 0.0044 & 2.9002 & 247.49 & -0.369973 & 6.800018  & 13.456769 & 6.559029 \\
2455836.50488464 & 0.08431 & 0.00046 & 7.08978 & -0.02477 & -4.9450 & 0.0027 & 2.9396 & 171.93 & -0.148130 & 6.458323  & 13.693823 & 6.750327 \\
2455836.63463460 & 0.10254 & 0.00047 & 7.08721 & -0.02589 & -4.9506 & 0.0029 & 2.9487 & 19.69  & -0.149324 & 6.451065  & 13.612058 & 6.682767 \\
2455837.58225198 & 0.09916 & 0.00043 & 7.08632 & -0.02437 & -4.9575 & 0.0026 & 2.9941 & 112.57 & -0.062027 & 7.066728  & 13.760690 & 6.785311 \\
2455837.66801992 & 0.10148 & 0.00054 & 7.08911 & -0.02632 & -4.9597 & 0.0041 & 3.0056 & 252.95 & -0.062027 & 7.066728  & 13.760690 & 6.785311 \\
2455838.56557661 & 0.09552 & 0.00065 & 7.09328 & -0.02698 & -4.9373 & 0.0043 & 3.0525 & 261.60 & -0.104478 & 7.448517  & 13.797603 & 6.798914 \\
2455838.67384272 & 0.09925 & 0.00077 & 7.0939  & -0.02717 & -4.9475 & 0.0061 & 3.0623 & 81.28  & -0.104478 & 7.448517  & 13.797603 & 6.798914 \\
2455839.56369186 & 0.10600 & 0.00049 & 7.09053 & -0.02331 & -4.9480 & 0.0031 & 3.1099 & 78.70  & 0.282247  & 6.290580  & 13.538572 & 6.711449 \\
2455839.66500316 & 0.10144 & 0.00054 & 7.08958 & -0.02547 & -4.9404 & 0.0038 & 3.1058 & 241.53 & 0.282247  & 6.290580  & 13.538572 & 6.711449 \\
2455840.54096382 & 0.10181 & 0.00042 & 7.09045 & -0.02573 & -4.9519 & 0.0024 & 3.1535 & 216.48 & 0.492760  & 6.355797  & 13.337171 & 6.545630 \\
2455841.51821591 & 0.09743 & 0.00044 & 7.0908  & -0.02434 & -4.9443 & 0.0023 & 3.2110 & 359.87 & 0.548444  & 6.054414  & 13.267717 & 6.419010 \\
2455842.54067654 & 0.10156 & 0.00058 & 7.09398 & -0.02331 & -4.9419 & 0.0041 & 3.2579 & 210.63 & 0.285453  & 5.964030  & 13.453311 & 6.502014 \\
2455842.64940682 & 0.10221 & 0.00065 & 7.09629 & -0.02221 & -4.9521 & 0.0051 & 3.2673 & 30.31  & 0.285453  & 5.964030  & 13.453311 & 6.502014 \\
2455843.53446953 & 0.10135 & 0.00042 & 7.09755 & -0.02479 & -4.9529 & 0.0023 & 3.3128 & 16.47  & 0.289295  & 6.042467  & 14.255929 & 7.096244 \\
2455843.66473254 & 0.09722 & 0.00057 & 7.09789 & -0.0264  & -4.9598 & 0.0043 & 3.3214 & 229.83 & 0.289295  & 6.042467  & 14.255929 & 7.096244 \\
2455844.53056656 & 0.09656 & 0.00058 & 7.09006 & -0.02479 & -4.9538 & 0.0039 & 3.3722 & 187.91 & 0.338963  & 7.427578  & 15.036871 & 7.600587 \\
2455844.66179197 & 0.10292 & 0.00050 & 7.09519 & -0.0258  & -4.9513 & 0.0033 & 3.3799 & 41.27  & 0.412640  & 7.942586  & 15.167705 & 7.742510 \\
2455845.54569082 & 0.10143 & 0.00053 & 7.09354 & -0.0214  & -4.9472 & 0.0032 & 3.4265 & 33.04  & 0.214037  & 9.072355  & 15.656077 & 8.091010 \\
2455845.65206487 & 0.09830 & 0.00054 & 7.09719 & -0.02343 & -4.9456 & 0.0036 & 3.4313 & 201.48 & 0.147598  & 9.104808  & 15.530288 & 7.938438 \\
2455870.51420322 & 0.10843 & 0.00059 & 7.09849 & -0.01902 & -4.9355 & 0.0032 & 4.7844 & 86.70  & -0.230017 & 9.870595  & 18.187723 & 9.696133 \\
2455870.52215421 & 0.10701 & 0.00053 & 7.10107 & -0.02157 & -4.9399 & 0.0035 & 4.7834 & 97.93  & -0.230017 & 9.870595  & 18.187723 & 9.696133 \\
2455870.53060607 & 0.10379 & 0.00053 & 7.09826 & -0.02232 & -4.9419 & 0.0036 & 4.7819 & 109.16 & -0.230017 & 9.870595  & 18.187723 & 9.696133 \\
2455871.50280525 & 0.10250 & 0.00061 & 7.09361 & -0.01883 & -4.9341 & 0.0033 & 4.8375 & 241.14 & -0.247643 & 9.628828  & 17.731211 & 9.520507 \\
2455871.51026399 & 0.10190 & 0.00053 & 7.09764 & -0.02095 & -4.9334 & 0.0033 & 4.8370 & 252.37 & -0.247643 & 9.628828  & 17.731211 & 9.520507 \\
2455871.51752074 & 0.10279 & 0.00053 & 7.09829 & -0.02004 & -4.9307 & 0.0033 & 4.8366 & 269.21 & -0.247643 & 9.628828  & 17.731211 & 9.520507 \\
2455872.49503353 & 0.10638 & 0.00177 & 7.10366 & -0.02363 & -4.8856 & 0.0193 & 4.8869 & 46.81  & -0.217090 & 8.539591  & 16.708896 & 8.931849 \\
2455872.50361680 & 0.10706 & 0.00063 & 7.09857 & -0.0206  & -4.9459 & 0.0044 & 4.8853 & 63.65  & -0.212479 & 8.339659  & 16.718279 & 8.928224 \\
2455872.51065080 & 0.10433 & 0.00066 & 7.09374 & -0.02182 & -4.9417 & 0.0046 & 4.8852 & 74.88  & -0.232778 & 8.259415  & 16.704289 & 8.916693 \\
2455872.51832551 & 0.10431 & 0.00066 & 7.09474 & -0.02018 & -4.9297 & 0.0044 & 4.8914 & 86.10  & -0.218351 & 8.044784  & 16.715321 & 8.926018 \\
2455873.50406872 & 0.09870 & 0.00050 & 7.09124 & -0.02293 & -4.9400 & 0.0030 & 4.9404 & 240.54 & -0.159657 & 7.999471  & 15.740116 & 8.388542 \\
2455873.51131777 & 0.09937 & 0.00050 & 7.09252 & -0.02197 & -4.9425 & 0.0031 & 4.9401 & 251.76 & -0.149055 & 8.115433  & 15.749984 & 8.389838 \\
2455873.51863681 & 0.09933 & 0.00051 & 7.09635 & -0.02245 & -4.9380 & 0.0030 & 4.9397 & 262.99 & -0.137447 & 8.065520  & 15.735783 & 8.370678 \\
2455874.50355099 & 0.10133 & 0.00077 & 7.09652 & -0.02386 & -4.9467 & 0.0051 & 4.9964 & 57.42  & -0.289008 & 8.449905  & 15.011817 & 7.870854 \\
2455874.51074342 & 0.10294 & 0.00074 & 7.09261 & -0.02314 & -4.9402 & 0.0057 & 4.9962 & 68.65  & -0.305588 & 8.072379  & 14.993994 & 7.858010 \\
2455874.51805876 & 0.10132 & 0.00089 & 7.09796 & -0.02544 & -4.9379 & 0.0069 & 4.9958 & 79.87  & -0.305588 & 8.072379  & 14.993994 & 7.858010 \\
2455875.50408302 & 0.09839 & 0.00058 & 7.08979 & -0.02418 & -4.9544 & 0.0040 & 5.0515 & 234.30 & -0.160578 & 7.051506  & 14.539260 & 7.545515 \\
2455875.51155021 & 0.09671 & 0.00057 & 7.09151 & -0.02384 & -4.9472 & 0.0039 & 5.0509 & 245.52 & -0.160578 & 7.051506  & 14.539260 & 7.545515 \\
2455875.51886866 & 0.09645 & 0.00058 & 7.09236 & -0.01925 & -4.9500 & 0.0040 & 5.0506 & 256.75 & -0.160578 & 7.051506  & 14.539260 & 7.545515 \\
2455878.50606288 & 0.10481 & 0.00123 & 7.09306 & -0.02384 & -4.9368 & 0.0129 & 5.2162 & 44.91  & 0.239367  & 7.862843  & 14.083289 & 7.241949 \\
2455878.51386630 & 0.10396 & 0.00125 & 7.08715 & -0.02017 & -4.9521 & 0.0137 & 5.2153 & 56.14  & 0.219175  & 7.727767  & 14.087698 & 7.262066 \\
2455878.52076438 & 0.10324 & 0.00120 & 7.09272 & -0.02694 & -4.9529 & 0.0128 & 5.2153 & 67.36  & 0.215755  & 7.792477  & 14.102572 & 7.257334 \\
2455879.50702795 & 0.09725 & 0.00112 & 7.09457 & -0.02522 & -4.9197 & 0.0103 & 5.2777 & 227.38 & 0.232115  & 7.128572  & 14.466632 & 7.447019 \\
2455879.51427708 & 0.09705 & 0.00107 & 7.08951 & -0.02834 & -4.9360 & 0.0100 & 5.2774 & 238.61 & 0.230304  & 6.983342  & 14.483479 & 7.457781 \\
2455879.52152229 & 0.09781 & 0.00105 & 7.09496 & -0.02536 & -4.9146 & 0.0093 & 5.2771 & 249.84 & 0.232562  & 7.438171  & 14.526323 & 7.492319 \\
2455886.55971131 & 0.09668 & 0.00085 & 7.09998 & -0.02119 & -4.9537 & 0.0073 & 5.6625 & 109.52 & 0.232784  & 6.416208  & 14.173657 & 6.983742 \\
2455886.56696080 & 0.09496 & 0.00086 & 7.09483 & -0.02348 & -4.9481 & 0.0075 & 5.6622 & 120.75 & 0.232784  & 6.416208  & 14.173657 & 6.983742 \\
2455886.57392844 & 0.09257 & 0.00093 & 7.10074 & -0.0237  & -4.9477 & 0.0086 & 5.6622 & 131.98 & 0.232784  & 6.416208  & 14.173657 & 6.983742 \\
2455887.52342644 & 0.09687 & 0.00060 & 7.09076 & -0.02392 & -4.9636 & 0.0042 & 5.7196 & 224.60 & 0.254368  & 6.578471  & 13.681434 & 6.611017 \\
2455887.53095415 & 0.09745 & 0.00067 & 7.09191 & -0.02685 & -4.9726 & 0.0052 & 5.7190 & 235.83 & 0.254368  & 6.578471  & 13.681434 & 6.611017 \\
2455887.53806579 & 0.09616 & 0.00072 & 7.09353 & -0.02617 & -4.9578 & 0.0056 & 5.7189 & 247.06 & 0.254368  & 6.578471  & 13.681434 & 6.611017 \\
2455888.52593242 & 0.10110 & 0.00060 & 7.09039 & -0.02613 & -4.9661 & 0.0043 & 5.7727 & 47.04  & 0.157666  & 6.316140  & 13.525647 & 6.498637 \\
2455888.53317611 & 0.10158 & 0.00059 & 7.08831 & -0.02224 & -4.9623 & 0.0043 & 5.7724 & 58.27  & 0.157666  & 6.316140  & 13.525647 & 6.498637 \\
2455888.54049630 & 0.10175 & 0.00062 & 7.09083 & -0.02548 & -4.9577 & 0.0047 & 5.7720 & 69.50  & 0.157666  & 6.316140  & 13.525647 & 6.498637

\end{tabular}}
\end{center}
\end{table*}

\begin{table*}
\contcaption{}
\begin{center}

{\scriptsize
\begin{tabular}{cccccccccccccc}
Julian Date & RV & $\sigma_{\rm RV}$ & FWHM & BIS & $\log(R'_{\rm HK}$) & $\sigma_{\log(R'_{\rm HK})}$ & $\Delta t$ & $\lambda$ & $ \hat{v}_{{\rm phot}}$ & $ \hat{v}_{{\rm conv}}$ & $|\hat{B}_{\rm obs}|$ & $f$\\

[MJD$_{\rm mid}$\_UTC] & [km~s$^{-1}$] & [km~s$^{-1}$] & [km~s$^{-1}$] & [km~s$^{-1}$] &    &   & [days] & [deg] & [m~s$^{-1}$] & [m~s$^{-1}$] & [G] & [\%] \\
\hline
\rule{0pt}{0ex} \\

2455889.52578074 & 0.09755 & 0.00060 & 7.09131 & -0.02611 & -4.9642 & 0.0045 & 5.8284 & 223.87 & 0.183892  & 6.607664  & 13.455927 & 6.430930 \\
2455889.53310116 & 0.09601 & 0.00060 & 7.09305 & -0.02411 & -4.9633 & 0.0046 & 5.8280 & 235.10 & 0.183892  & 6.607664  & 13.455927 & 6.430930 \\
2455889.54034195 & 0.09461 & 0.00065 & 7.09377 & -0.02242 & -4.9796 & 0.0054 & 5.8277 & 246.32 & 0.183892  & 6.607664  & 13.455927 & 6.430930 \\
2455890.50917526 & 0.10097 & 0.00059 & 7.08886 & -0.02867 & -4.9505 & 0.0040 & 5.8867 & 12.62  & 0.079914  & 6.820067  & 13.946109 & 6.891653 \\
2455890.51649149 & 0.10074 & 0.00058 & 7.08763 & -0.02659 & -4.9564 & 0.0040 & 5.8863 & 23.85  & 0.079914  & 6.820067  & 13.946109 & 6.891653 \\
2455890.52382222 & 0.09996 & 0.00059 & 7.08779 & -0.02485 & -4.9611 & 0.0042 & 5.8859 & 35.08  & 0.079914  & 6.820067  & 13.946109 & 6.891653 \\
2455891.52246765 & 0.09941 & 0.00060 & 7.09151 & -0.02535 & -4.9591 & 0.0044 & 5.9428 & 211.90 & 0.079914  & 6.820067  & 13.946109 & 6.891653 \\
2455891.52978471 & 0.09921 & 0.00063 & 7.0924  & -0.02549 & -4.9649 & 0.0049 & 5.9424 & 223.12 & -0.128128 & 7.724229  & 15.031973 & 7.761893 \\
2455891.53703495 & 0.09840 & 0.00063 & 7.09353 & -0.02616 & -4.9482 & 0.0049 & 5.9421 & 234.35 & -0.128128 & 7.724229  & 15.031973 & 7.761893 \\
2455892.52735539 & 0.10398 & 0.00061 & 7.08642 & -0.02254 & -4.9442 & 0.0045 & 5.9935 & 34.33  & 0.250746  & 8.037871  & 15.638486 & 7.970436 \\
2455892.53453612 & 0.10358 & 0.00079 & 7.09145 & -0.02441 & -4.9370 & 0.0068 & 5.9932 & 45.55  & 0.239745  & 8.154407  & 15.602640 & 7.956911 \\
2455892.54214391 & 0.10247 & 0.00069 & 7.0878  & -0.02379 & -4.9422 & 0.0054 & 5.9995 & 62.40  & 0.239745  & 8.154407  & 15.602640 & 7.956911 \\
2455893.51437423 & 0.10022 & 0.00059 & 7.09516 & -0.02493 & -4.9440 & 0.0040 & 6.0551 & 194.30 & 0.178788  & 9.501673  & 16.492246 & 8.578127 \\
2455893.52626540 & 0.10239 & 0.00057 & 7.09183 & -0.02514 & -4.9578 & 0.0040 & 6.0501 & 211.14 & 0.178788  & 9.501673  & 16.492246 & 8.578127 \\
2455893.53359078 & 0.10228 & 0.00060 & 7.09683 & -0.02459 & -4.9607 & 0.0044 & 6.0497 & 222.37 & 0.178788  & 9.501673  & 16.492246 & 8.578127 \\
2455894.50541095 & 0.10265 & 0.00059 & 7.09496 & -0.02371 & -4.9499 & 0.0041 & 6.0988 & 354.27 & -0.143119 & 10.188700 & 16.843670 & 8.978122 \\
2455894.51266019 & 0.10400 & 0.00062 & 7.09307 & -0.02355 & -4.9416 & 0.0044 & 6.1054 & 5.49   & -0.143119 & 10.188700 & 16.843670 & 8.978122 \\
2455894.52011844 & 0.10548 & 0.00068 & 7.09466 & -0.02304 & -4.9486 & 0.0052 & 6.1049 & 16.72  & -0.143119 & 10.188700 & 16.843670 & 8.978122 \\
2455895.51190055 & 0.09855 & 0.00058 & 7.0944  & -0.0217  & -4.9341 & 0.0038 & 6.1617 & 182.30 & -0.219694 & 9.459038  & 16.678771 & 8.985445 \\
2455895.51928533 & 0.10014 & 0.00061 & 7.09307 & -0.02248 & -4.9339 & 0.0041 & 6.1613 & 193.53 & -0.219694 & 9.459038  & 16.678771 & 8.985445 \\
2455895.52668202 & 0.10385 & 0.00060 & 7.09409 & -0.02013 & -4.9456 & 0.0041 & 6.1608 & 204.76 & -0.219694 & 9.459038  & 16.678771 & 8.985445 \\
2455896.51221054 & 0.10311 & 0.00062 & 7.09555 & -0.0246  & -4.9082 & 0.0042 & 6.2170 & 359.11 & -0.118240 & 9.149870  & 16.324338 & 8.893371 \\
2455896.51945075 & 0.10402 & 0.00064 & 7.0943  & -0.02161 & -4.9219 & 0.0044 & 6.2167 & 10.34  & -0.118240 & 9.149870  & 16.324338 & 8.893371 \\
2455896.52676887 & 0.10450 & 0.00065 & 7.0945  & -0.0215  & -4.9086 & 0.0045 & 6.2163 & 21.56  & -0.118240 & 9.149870  & 16.324338 & 8.893371 \\
2455897.51313557 & 0.09805 & 0.00057 & 7.0929  & -0.02141 & -4.9279 & 0.0041 & 6.2716 & 175.91 & 0.310313  & 8.521387  & 15.731010 & 8.518003 \\
2455897.52017185 & 0.10199 & 0.00061 & 7.09074 & -0.02029 & -4.9207 & 0.0045 & 6.2715 & 187.14 & 0.310313  & 8.521387  & 15.731010 & 8.518003 \\
2455897.52797234 & 0.10103 & 0.00066 & 7.09431 & -0.02293 & -4.9178 & 0.0049 & 6.2776 & 203.98 & 0.310313  & 8.521387  & 15.731010 & 8.518003 \\
2455898.51251923 & 0.10034 & 0.00076 & 7.09442 & -0.02529 & -4.9139 & 0.0059 & 6.3278 & 352.71 & 0.208771  & 7.346064  & 15.023078 & 7.994100 \\
2455898.51955262 & 0.10292 & 0.00074 & 7.09572 & -0.02088 & -4.9290 & 0.0060 & 6.3277 & 3.94   & 0.193525  & 7.375161  & 15.016566 & 7.987264 \\
2455898.52692661 & 0.10153 & 0.00078 & 7.09194 & -0.01818 & -4.9312 & 0.0067 & 6.3272 & 15.17  & 0.193006  & 7.321089  & 15.020403 & 7.988140 \\
2455899.51280365 & 0.09428 & 0.00077 & 7.09221 & -0.02172 & -4.9295 & 0.0065 & 6.3830 & 169.51 & 0.355292  & 6.343507  & 14.470741 & 7.472408 \\
2455899.51984723 & 0.09535 & 0.00075 & 7.08803 & -0.02529 & -4.9478 & 0.0066 & 6.3829 & 180.74 & 0.360379  & 6.548910  & 14.477587 & 7.468980 \\
2455899.52750987 & 0.09653 & 0.00076 & 7.09312 & -0.02194 & -4.9296 & 0.0063 & 6.3822 & 191.96 & 0.340881  & 6.579109  & 14.478097 & 7.470433 \\
2455900.51316244 & 0.09920 & 0.00071 & 7.08969 & -0.02418 & -4.9457 & 0.0060 & 6.4382 & 346.30 & 0.270016  & 6.616161  & 13.979841 & 7.170614 \\
2455900.52048381 & 0.09737 & 0.00071 & 7.09126 & -0.02617 & -4.9389 & 0.0059 & 6.4378 & 357.53 & 0.290577  & 6.656814  & 13.981226 & 7.170553 \\
2455900.52779857 & 0.09982 & 0.00073 & 7.0925  & -0.02389 & -4.9388 & 0.0062 & 6.4444 & 14.37  & 0.287024  & 6.723330  & 13.985253 & 7.170031 \\
2455901.51376528 & 0.09412 & 0.00084 & 7.08755 & -0.02695 & -4.9244 & 0.0074 & 6.4932 & 163.10 & 0.527470  & 5.870234  & 13.607637 & 6.908622 \\
2455901.52080259 & 0.09432 & 0.00079 & 7.09411 & -0.02379 & -4.9370 & 0.0068 & 6.4931 & 174.32 & 0.526638  & 5.968377  & 13.626008 & 6.919914 \\
2455901.52818692 & 0.09743 & 0.00084 & 7.08943 & -0.01941 & -4.9387 & 0.0078 & 6.4996 & 191.16 & 0.528856  & 5.908731  & 13.682738 & 6.957918 \\
2455902.51342158 & 0.09849 & 0.00100 & 7.09581 & -0.02293 & -4.9453 & 0.0099 & 6.5491 & 339.89 & 0.080486  & 5.991561  & 13.461406 & 6.723588 \\
2455902.52087765 & 0.09811 & 0.00094 & 7.08712 & -0.02349 & -4.9608 & 0.0097 & 6.5555 & 356.73 & 0.072490  & 6.135274  & 13.455276 & 6.727012 \\
2455902.52825791 & 0.09928 & 0.00096 & 7.08618 & -0.02403 & -4.9391 & 0.0096 & 6.5551 & 7.96   & 0.084431  & 6.272675  & 13.472634 & 6.750239

\end{tabular}}
\end{center}
\end{table*}

\bsp	
\label{lastpage}
\end{document}